\documentclass[aps,prd,twoside,twocolumn,nofootinbib,10pt,showpacs,floatfix]{revtex4-1}
\usepackage{graphicx,bm}
\usepackage{slashed}
\usepackage{epstopdf}
\usepackage{ulem} 
\usepackage[usenames]{color}
\usepackage{float}
\usepackage{hyperref}
\usepackage{subfigure}
\usepackage{subfigure}
\usepackage{rotating}
\usepackage{color}
\usepackage{multirow}
\usepackage{dcolumn}
\usepackage{orcidlink}
\usepackage{overpic}
\usepackage{booktabs}
\usepackage{makecell}
\usepackage{amsmath,amssymb,bm}
\usepackage{array}

\newcommand{\TpsiZero}{T_{\psi0}^{a}(4020)}
\newcommand{\TpsiTwo}{T_{\psi2}^{a}(4020)}

\newcommand{\dd}{\mathrm{d}}
\newcommand{\ii}{\mathrm{i}}
\newcommand{\MeV}{\mathrm{MeV}}
\newcommand{\GeV}{\mathrm{GeV}}

\begin{document}

\title{Implications of
$B\to \chi_{c1}\pi K$ data for an isovector
$G$-odd $D^\ast\bar D^\ast$ molecular virtual state}

\author{Jun-Zhang Wang \orcidlink{0000-0002-3404-8569}}\email{wangjzh@cqu.edu.cn}
\affiliation{Department of Physics and Chongqing Key Laboratory for Strongly Coupled Physics, Chongqing University, Chongqing 401331, China}

\date{\today}

\begin{abstract}
Establishing the near-threshold spin-isospin multiplet spectrum of
$D^{(\ast)}\bar D^{(\ast)}$ systems is central to testing molecular
interpretations of the $X(3872)$, $Z_c(3900)$, and $Z_c(4020)$.
The isovector channels are particularly important in this context, as the associated structures would have a clear exotic character.
In this work, we analyze the BaBar and Belle data on
$B\to\chi_{c1}\pi K$ to search for possible isovector $D^\ast\bar D^\ast$ molecules with $J^{PC}=0^{++}$ and $2^{++}$.  The three-body decay
amplitude includes an effective nonresonant term, intermediate kaon
resonances, and
$\chi_{c1}\pi$--$D^\ast\bar D^\ast$ coupled-channel rescattering.  For each data set, the $\chi_{c1}\pi$ and $K\pi$ invariant-mass distributions are fitted simultaneously under three scenarios: without rescattering and with either $J=0$ or $J=2$ rescattering. We then
analytically continue the fitted coupled-channel $T$ matrices to
search for poles.  For the case of $J^{PC}=0^{++}$, the BaBar and
Belle fits yield virtual poles at $-2.15^{+2.10}_{-6.41}~\MeV$ and
$-18.40^{+8.58}_{-13.70}~\MeV$, respectively, relative to the
$D^\ast\bar D^\ast$ threshold.  For $J^{PC}=2^{++}$, the corresponding virtual poles are located at $-3.42^{+2.37}_{-4.29}~\MeV$ and
$-4.26^{+1.34}_{-1.69}~\MeV$, respectively.  The tensor virtual-state pole is consistent with a prediction from chiral
effective field theory, which also predicts the existence of $W_{c1}$, an isospin partner of $X(3872)$.  Since the invariant-mass distributions alone cannot distinguish total spin $J$, we also predict angular distributions in the three-body final state.  The
$\cos\theta_{\chi_{c1}\pi}$ distribution provides a direct spin
discriminator, while the
$\cos\theta_{K\pi}$ distribution further tests the coupled-channel rescattering contribution.
Measurements of these predicted distributions by the Belle II and LHCb collaborations would help establish the $D^{(\ast)}\bar D^{(\ast)}$ molecular multiplet spectrum in the future.
\end{abstract}

\maketitle

\section{Introduction}
\label{sec:introduction}

Over the past two decades, numerous charmoniumlike structures have been observed near open-charm thresholds, making hadronic molecules an important topic in hidden-charm $XYZ$ spectroscopy~\cite{Chen:2016qju,Hosaka:2016pey,Lebed:2016hpi,Esposito:2016noz,Olsen:2017bmm,Guo:2017jvc,Liu:2019zoy,Brambilla:2019esw,Meng:2022ozq,Liu:2024uxn,Bai:2026atm,Dai:2026fkg,Wang:2025dur,Wang:2025sic}.  The $X(3872)$, first observed by Belle~\cite{Belle:2003nnu}, lies at the $D^0\bar D^{\ast0}$ threshold and has quantum numbers $J^{PC}=1^{++}$~\cite{LHCb:2013kgk}. Its decays into $J/\psi\rho^0$ and $J/\psi\omega$ exhibit pronounced isospin breaking~\cite{Guo:2017jvc}.  The BESIII Collaboration observed the charged
$Z_c(3900)$ and $Z_c(4020)$ close to the $D\bar D^\ast$
and $D^\ast\bar D^\ast$ thresholds, showing that the isovector sector also
contains prominent near-threshold hidden-charm structures
\cite{BESIII:2013ris,BESIII:2013ouc,BESIII:2013mhi}.  If these structures arise from charmed-meson anticharmed-meson interactions, heavy-quark spin symmetry relates different spin channels and may imply additional near-threshold partners. Establishing the resulting S-wave multiplet therefore provides a key test of the molecular interpretation.

The neutral partners of the observed $Z_c(3900)$ and $Z_c(4020)$ have $J^{PC}=1^{+-}$ and, in molecular interpretations~\cite{Guo:2013sya,Dong:2013iqa,He:2013nwa,Ke:2013gia,Chen:2013omd,Aceti:2014uea,Albaladejo:2015lob,Gong:2016hlt,Pilloni:2016obd,Ortega:2018cnm,Du:2020vwb,Yang:2020nrt,Baru:2021ddn,Chen:2022ddj,Chen:2023def,Wilbring:2013cha,Wang:2013cya,Lin:2024qcq,Aceti:2014kja,Wu:2023rrp,Liu:2024ziu,Yu:2024sqv}, are associated with S-wave $D\bar D^\ast$ and $D^\ast\bar D^\ast$ configurations, respectively.  Here we focus on the distinct
isovector S-wave channels with $J^{PC}=1^{++}$ for $D\bar D^\ast$ and
$J^{PC}=0^{++},2^{++}$ for $D^\ast\bar D^\ast$.  Throughout this work,
$J^{PC}$ labels the neutral member of an isovector multiplet and is also used to label its charged partners.\footnote{For an isovector multiplet, $G=C(-1)^I=-C$, so channels whose neutral members have $C=+$ are $G$-odd.  The $J^{PC}$ notation used below should not be read as assigning $C$ parity to a charged
state.}
At leading order in a heavy-meson contact effective field theory (EFT),
heavy-quark spin symmetry
gives
\[
  V(1^{++},D\bar D^\ast)=C_{1a}+C_{1b},\quad
  V(2^{++},D^\ast\bar D^\ast)=C_{1a}+C_{1b},
\]
while the scalar $D^\ast\bar D^\ast$ channel is governed by
\[
  V(0^{++},D^\ast\bar D^\ast)=C_{1a}-2C_{1b}.
\]
Here $C_{1a}$ and $C_{1b}$ are low-energy constants in the isovector
contact potential.  The equality of the $1^{++}$ and $2^{++}$ potentials
connects the tensor $D^\ast\bar D^\ast$ system to the isovector
$D\bar D^\ast$ interaction, whereas the scalar $D^\ast\bar D^\ast$
system depends on a different short-range combination
\cite{Hidalgo-Duque:2012rqv,Baru:2021ddn,Zhang:2024fxy}.

Table~\ref{tab:theory-status} summarizes representative predictions for
these isovector channels.  One-boson-exchange (OBE) models find the
isovector attraction weakened by the flavor factors of the exchanged
mesons.  With natural cutoffs, the OBE models generally tend to disfavor bound states,
particularly in the $2^{++}$ channel, while often finding stronger
attraction in the $0^{++}$ channel
\cite{Liu:2009qhy,He:2013nwa,Liu:2019stu,Ding:2020dio}. 
Contact EFT calculations give a much wider
range of both scalar and tensor poles, from deeply bound to near-threshold bound, virtual,
or resonant states, and in some cases no pole.  This spread reflects the different inputs used to constrain the low-energy
coupling constants.  We denote the predicted isovector scalar and tensor pole states by $\TpsiZero$ and $\TpsiTwo$, respectively, with the superscript $a$ indicating an isovector multiplet~\cite{Gershon:2022xnn}.

\begin{table*}[t]
\caption{
Representative pole predictions for the three $S$-wave isovector molecular
partners. The first column specifies the dynamical framework and, where relevant, the input or fit used to constrain it or the numerical method employed to determine the pole positions.  Here $X\equiv X(3872)$. 
The contact EFT entries are
leading-order calculations.  
Quasipotential Bethe--Salpeter equation (qBSE) and complex-scaling
method (CSM) are abbreviated in the first column.  Pole positions are in MeV, and
$\text{B}$, $\text{V}$, and $\text{R}$ denote bound, virtual, and
resonance poles.  ``No bound'' indicates a search restricted to bound states,
and ``$\cdots$'' corresponds to an unstudied channel.  Threshold-relative results are converted
using the specified threshold,
$M_{\rm th}(D\bar D^\ast)=3875.8~\MeV$,
$M_{\rm th}(D^+D^{\ast-})=3879.9~\MeV$, and
$M_{\rm th}(D^\ast\bar D^\ast)=4018.0~\MeV$. If absolute pole positions are available in the original references, they are used directly here. Distinct
solutions from the same study are listed separately.
}
\label{tab:theory-status}
\centering
\begingroup
\scriptsize
\setlength{\tabcolsep}{3pt}
\renewcommand{\arraystretch}{1.26}
\newcommand{\nopole}{$\text{no pole}$}
\newcommand{\nobound}{$\text{no bound}$}
\newcommand{\notstudied}{$\cdots$}
\begin{ruledtabular}
\begin{tabular}{lccc}
Model: input/method
& $W_{c1}$, $D\bar D^\ast$, $1^{++}$
& $T_{\psi0}^{a}$, $D^\ast\bar D^\ast$, $0^{++}$
& $T_{\psi2}^{a}$, $D^\ast\bar D^\ast$, $2^{++}$ \\
\colrule
Contact EFT: $X/X(3915)/Y(4140)$ \cite{Hidalgo-Duque:2012rqv}
& \nopole
& $3960^{+31}_{-37}\;(\Lambda=1~\GeV),\ \text{B}$
& \nopole \\

Contact EFT: $S$-wave production fit \cite{Du:2020vwb}
& $3869.09,\ \text{B}$
& \nopole
& $4011.14,\ \text{B}$ \\

Contact EFT: $D$-wave production fit \cite{Du:2020vwb}
& \nopole
& $4022.07-i\,6.58,\ \text{R}$
& \nopole \\

Contact EFT: $Z_c/Z_{cs}$ fit 1 \cite{Baru:2021ddn}
& $3864\pm2,\ \text{V}$
& \notstudied
& $4009\pm2,\ \text{V}$ \\

Contact EFT: $Z_c/Z_{cs}$ fit 2 \cite{Baru:2021ddn}
& $\simeq3852,\ \text{B}$
& \notstudied
& $\simeq3990,\ \text{B}$ \\

Contact EFT: $X/Z_c/X(3960)$ \cite{Ji:2022uie}
& \nopole
& $3972.1^{+30}_{-56}\;(\Lambda=1~\GeV),\ \text{V}$
& $3944.1^{+46}_{-76}\;(\Lambda=0.5~\GeV),\ \text{B}$ \\

Contact EFT: saturated meson exchange 
\cite{Peng:2023lfw}
& \nobound
& $4004.9^{+5.6}_{-8.5},\ \text{B}$
& \nopole \\

Contact EFT: light-quark interaction 
\cite{Wang:2023hpp}
& \nopole
& $4007.2\text{--}4016.7,\ \text{B}$
& \nopole \\

Chiral EFT: $X$ input \cite{Zhang:2024fxy}
& $\begin{array}{c}
W_{c1}^{0}:\ 3881.2^{+0.8}_{-0.0}-i\,1.6^{+0.7}_{-0.9},\ \text{V}\\
W_{c1}^{\pm}:\ 3866.9^{+4.6}_{-7.7}-i\,0.07,\ \text{V}
\end{array}$
& \notstudied
& $4010.4^{+2.2}_{-3.5},\ \text{V}$ \\

Chiral EFT: dispersive analysis of $X$ decays \cite{Dias:2024zfh}
& $\begin{array}{c}
W_{c1}^{0}:\ 3881.7^{+1.0}_{-0.7}+i\,1.2^{+0.8}_{-0.7},\ \text{V}\\
W_{c1}^{\pm}:\ 3862.5^{+6.4}_{-10.3}-i\,0.07,\ \text{V}
\end{array}$
& \notstudied
& \notstudied \\

Chiral EFT: pole fit of $X$ \cite{Ji:2025hjw}
& $W_{c1}^{0}:\ 3883.0\pm0.7+i\,1.3^{+1.9}_{-0.6},\ \text{V}$
& \notstudied
& \notstudied \\

OBE model \cite{Liu:2009qhy}
& \nobound
& \nobound
& \nobound \\

OPE model \cite{He:2013nwa}
& \notstudied
& $\simeq4025,\ \text{B}$
& \nobound \\

OBE model: $X$ input \cite{Liu:2019stu}
& \nobound
& \nobound
& \nobound \\

OBE model: qBSE \cite{Ding:2020dio}
& \nobound
& $4016.1\;(\Lambda=1.8~\GeV),\ \text{B}$
& \nobound \\

OBE model: CSM \cite{Lu:2025zae}
& $3875.4\;(\Lambda=1.25~\GeV),\ \text{B}$
& $4017.8\;(\Lambda=1.25~\GeV),\ \text{B}$
& \nopole \\

OBE model: qBSE \cite{Chen:2025gxe}
& $\simeq3826\text{--}3876\;(\Lambda=1.8\text{--}3.5~\GeV),\ \text{V}$
& \notstudied
& \notstudied \\

Hidden-gauge model \cite{Molina:2009ct}
& \notstudied
& \nopole
& $3919-i\,74,\ \text{R}$ \\

Bethe--Salpeter model \cite{Li:2021jtq}
& \notstudied
& \nobound
& \notstudied \\
\end{tabular}
\end{ruledtabular}
\endgroup
\end{table*}

A comparatively reliable prediction with limited model dependence is provided by the chiral EFT analysis of Ref.~\cite{Zhang:2024fxy}. With both leading contact couplings fixed by the $X(3872)$ pole and its isospin-breaking decay-amplitude ratio into $J/\psi\rho^0$ and $J/\psi\omega$, and with one-pion exchange and $D\bar D\pi$ three-body dynamics treated explicitly, the analysis predicts neutral and charged $D\bar D^\ast$ virtual states, denoted by $W_{c1}^{0}$ and $W_{c1}^{\pm}$, with $W_{c1}^{0}$ being predominantly isovector with $J^{PC}=1^{++}$.
Within the same chiral EFT framework, refined $X(3872)$ inputs from a dispersive treatment of the $\pi\pi$ final-state interaction in $X(3872)\to J/\psi\pi\pi$~\cite{Dias:2024zfh} and a combined analysis of BESIII and LHCb data yield similar $W_{c1}$ virtual poles~\cite{Ji:2025hjw}. An independent lattice QCD calculation at $m_\pi\simeq280~\MeV$ also supports the existence of the $W_{c1}$~\cite{Sadl:2024dbd}.
Because heavy-quark spin symmetry assigns the same leading contact interaction to the isovector $1^{++}$ $D\bar D^\ast$ and $2^{++}$ $D^\ast\bar D^\ast$ channels, Ref.~\cite{Zhang:2024fxy} uses the interaction fixed in the $W_{c1}$ analysis to predict an isovector $2^{++}$ virtual pole at $4010.4^{+2.2}_{-3.5}~\MeV$ in a single-channel, pionless, isospin-symmetric calculation.  The neutral $W_{c1}^{0}$ signal is usually difficult to separate from the much stronger $X(3872)$ contribution~\cite{Zhang:2024fxy}, so the search for its $2^{++}$ isovector partner would provide an indirect test of the predicted $W_{c1}$ pole.

Experimentally, $B$ meson decays provide important access to these isovector
channels.  In the $\eta_c\pi$ invariant-mass distribution of
$B^0\to\eta_c(1S)K^+\pi^-$, in 2018, LHCb reported evidence for a charged hidden-charm tetraquark structure, denoted $Z_c(4100)^-$ or $X(4100)^-$
\cite{LHCb:2018oeg}, with $J^P=0^+$ and a significance of $3.4\sigma$.  
An S-wave
$\eta_c\pi$ can couple to the isovector $0^{++}$
$D^\ast\bar D^\ast$ channel, making this structure a natural scalar partner
candidate.
Moreover, in 2026,
LHCb used $9~{\rm fb}^{-1}$ of data and found a
satisfactory description of the $\eta_c \pi$ spectrum of $B \to \eta_c \pi K$ without a $Z_c(4100)^-$ contribution
\cite{LHCb:2025hbx}.  The present experimental status of $Z_c(4100)^-$ is therefore unsettled.

The $\chi_{c1}\pi$ channel serves as another important probe~\cite{Deng:2024pep,Nakamura:2019emd,Cao:2018vmv,Lee:2008gn}, because both the scalar and tensor $D^\ast\bar D^\ast$ channels can couple to it in $P$ wave, whereas the tensor contribution to $\eta_c\pi$ is suppressed by the higher partial wave.
In 2008, Belle studied $\bar B^0\to K^-\pi^+\chi_{c1}$ and reported two Breit-Wigner resonances in the $\pi^+\chi_{c1}$ spectrum, denoted $Z_1(4050)^+$ and $Z_2(4250)^+$~\cite{Belle:2008qeq}. The lower structure is close to the threshold of $D^*\bar D^*$. Its mass and width were determined to be
\[
M = 4051^{+14+20}_{-14-41}~\MeV,\qquad
\Gamma = 82^{+21+47}_{-17-22}~\MeV ,
\]
with a significance exceeding $5\sigma$ under model variations. 
In 2012, BaBar analyzed $B^0\to\chi_{c1}K^-\pi^+$ and $B^+\to\chi_{c1}K_S^0\pi^+$, modeling the reflections in the $\chi_{c1}\pi$ spectrum induced by the resonances observed in the $K\pi$ distribution, but found no evidence for either $Z_1(4050)^+$ or $Z_2(4250)^+$~\cite{BaBar:2011hrz}.
In 2016, Belle measured inclusive and exclusive $B$ decays to $\chi_{c1}$ and $\chi_{c2}$ using a high-statistics sample of $772\times10^6$ $B\bar B$ pairs, including the $B^0\to\chi_{c1}\pi^-K^+$ channel analyzed here, but did not perform an amplitude analysis of this mode~\cite{Belle:2015opn}. Therefore, an amplitude analysis of the new Belle data should help clarify whether an isovector state exists near the $D^\ast\bar D^\ast$ threshold.


In the BaBar analysis, this negative result was obtained by modeling a possible $Z_1(4050)^+$ contribution with a Breit-Wigner amplitude~\cite{BaBar:2011hrz}. Such a parameterization is not appropriate for a near-threshold hadronic molecule because it does not incorporate the $D^\ast\bar D^\ast$ threshold dynamics, and it also fails to satisfy unitarity~\cite{Guo:2017jvc}. A unitary  coupled-channel rescattering amplitude is therefore needed to further check this conclusion.

In this work, we implement this analysis with a coherent $B$ decay amplitude comprising an effective
nonresonant contribution, the intermediate $K^\ast(892)$,
$K_2^\ast(1430)$, and $K^\ast(1680)$ resonances, and a unitary
$\chi_{c1}\pi$--$D^\ast\bar D^\ast$ coupled-channel rescattering
contribution.  The BaBar and Belle data sets for $B\to\chi_{c1}\pi K$ are analyzed independently by simultaneously fitting the $\chi_{c1}\pi$ and $K\pi$ invariant-mass spectra in each data set. To assess the role of coupled-channel rescattering and compare the two $D^\ast\bar D^\ast$ spin assignments, we consider three fit schemes.
Scheme I contains the effective nonresonant and intermediate-kaon-resonance
contributions; Schemes II and III additionally include the $J=0$ and
$J=2$ coupled-channel rescattering contributions, respectively.  We then
analytically continue the coupled-channel $T$ matrices
to search for near-threshold poles in their different Riemann sheets.  We
also predict the $\cos\theta_{\chi_{c1}\pi}$ and
$\cos\theta_{K\pi}$ angular distributions in selected mass windows.  The former
discriminates the $J=0$ and $J=2$ hypotheses, while the latter further tests the role of the
threshold structure correlated with the rescattering contribution.

The paper is organized as follows.  Section~\ref{sec:framework} defines the
amplitude framework and the coupled-channel $T$ matrix.
Section~\ref{sec:invariant-mass} presents the invariant-mass fits and pole
analysis for the Belle and BaBar data.  Section~\ref{sec:angular} gives
angular-distribution predictions in selected $\chi_{c1}\pi$ and $K\pi$
mass windows.  The main conclusions are summarized in
Sec.~\ref{sec:summary}.

\section{Amplitude framework for
$B\to\chi_{c1}\pi K$}
\label{sec:framework}

\begin{figure*}[t]
\centering
\includegraphics[width=0.8\linewidth]{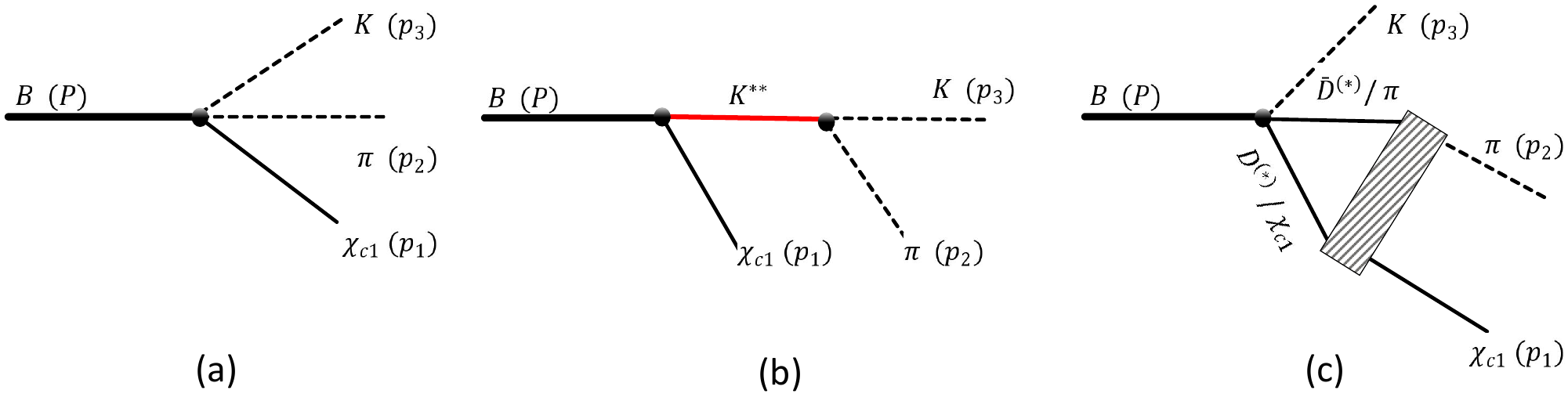}
\caption{
Decay mechanisms included in the amplitude: ($a$) effective nonresonant
three-body production, ($b$) $\chi_{c1}K^{\ast\ast}$ production followed by
$K^{\ast\ast}\to K\pi$, and ($c$) production of $K$ with either channel of the
$\chi_{c1}\pi$--$D^\ast\bar D^\ast$ system, followed by coupled-channel
rescattering.  The shaded block denotes the coupled-channel $T$ matrix.
}
\label{fig:decay-mechanism}
\end{figure*}

We consider the three-body decay
\begin{equation}
  B(P)\to \chi_{c1}(p_1,\epsilon_\chi)\,\pi(p_2)\,K(p_3),
  \label{eq:process}
\end{equation}
where $P$ is the $B$-meson four-momentum, $p_1,p_2,p_3$ are the final
$\chi_{c1},\pi,K$ four-momenta, and $\epsilon_\chi$ is the
$\chi_{c1}$ polarization vector.  The momentum flow follows
Fig.~\ref{fig:decay-mechanism}.  The decay
amplitude contains three classes of mechanisms: an effective nonresonant
three-body term, contributions from intermediate kaon meson resonances
decaying to $K\pi$, and a coupled-channel rescattering contribution in the
$\chi_{c1}\pi$--$D^\ast\bar D^\ast$ system.  The last mechanism is
described by a two-channel $T$ matrix.

The three two-body invariant masses are denoted by
\begin{equation}
  s_{12}=(p_1+p_2)^2,\qquad
  s_{23}=(p_2+p_3)^2,\qquad
  s_{13}=(p_1+p_3)^2 ,
  \label{eq:dalitz-variables}
\end{equation}
and satisfy
\begin{equation}
  s_{12}+s_{23}+s_{13}
  =
  m_B^2+m_{\chi_{c1}}^2+m_\pi^2+m_K^2 .
  \label{eq:dalitz-constraint}
\end{equation}
In what follows $m(\chi_{c1}\pi)=\sqrt{s_{12}}$ and
$m(K\pi)=\sqrt{s_{23}}$.

Figure~\ref{fig:decay-mechanism} also displays the weak production vertices used
in the decay amplitudes of different mechanisms.  In panel ($a$), the weak decay produces the
$\chi_{c1}\pi K$ final state without an intermediate two-body resonance, which defines the effective continuum amplitude.  We represent it by a
phenomenological effective three-body production vertex that includes direct
nonresonant production and the smooth background from nonresonant $K\pi$
interactions not included explicitly.  It is therefore distinct from the
production vertex for a $\chi_{c1}\pi$ pair in a specified partial wave in
panel ($c$).  In panel ($b$), the weak decay produces
$\chi_{c1}K^{\ast\ast}$, followed by $K^{\ast\ast}\to K\pi$. Here, $K^{\ast\ast}$ denotes the possible intermediate kaon mesonic states.  In panel ($c$), it
produces the bachelor $K$ together with either channel of the
$\chi_{c1}\pi$--$D^\ast\bar D^\ast$ system in a definite total angular
momentum $J$, and the two-body system then undergoes coupled-channel
rescattering. 

The effective coupling constants associated with the production vertices in these mechanisms are independent.  The effective continuum coupling $g_0$ is taken to be real.  The
kaon-resonance production couplings are
$g_1e^{\ii\phi_1}$, $g_2e^{\ii\phi_2}$, and $g_3e^{\ii\phi_3}$ for
$\chi_{c1}K^\ast(892)$, $\chi_{c1}K_2^\ast(1430)$, and
$\chi_{c1}K^\ast(1680)$, respectively.  Here $g_1,g_2,g_3$ are real
magnitudes and $\phi_1,\phi_2,\phi_3$ are phases relative to the effective
continuum amplitude.  For panel ($c$), $f_{Ja}$ and $f_{Jb}$ denote the
couplings for producing $K[\chi_{c1}\pi]_J$ and
$K[D^\ast\bar D^\ast]_J$, respectively, in the specified isovector
channel.  In particular, $f_{Jb}$ refers to the
$D^\ast\bar D^\ast$ pair projected onto the specified $S$-wave channel with $J=0$ or $J=2$. 
Since we focus on the invariant-mass line shapes, we retain only the leading Lorentz structures at the weak vertices displayed below, with the weak-decay dynamics absorbed into the effective production couplings introduced above.

\subsection{The conventional kaon-resonance mechanisms}

With these conventions, the effective continuum amplitude is taken in the
lowest-derivative form involving the final $\chi_{c1}$ polarization,
\begin{equation}
  \mathcal A_{\rm cont}
  =
  g_0\,\epsilon_\chi\cdot p_2 .
  \label{eq:continuum}
\end{equation}

We next specify the intermediate kaon-resonance contribution. In a three-body decay, an intermediate $K^{\ast\ast}$ band in the $K\pi$ invariant-mass distribution also produces a correlated structure in $m(\chi_{c1}\pi)$. We refer to this kinematic effect as a reflection. A reliable description of the kaon resonances in the $K\pi$ spectrum is therefore crucial for determining whether an observed structure in the $\chi_{c1}\pi$ spectrum is a reflection or has an independent dynamical origin.
The retained
states here are the vector $K^\ast(892)$, the tensor $K_2^\ast(1430)$, and the
vector radial excitation $K^\ast(1680)$, which are established members of the kaon meson spectrum. The measured $K\pi$ spectra have shown their resonant signals in the
corresponding mass regions, which also were included in the BaBar analysis of $B\to\chi_{c1}\pi K$~\cite{BaBar:2011hrz}.

For a vector $K^\ast\to K\pi$, we use the spin-one projector
\begin{equation}
  P_{\mu\nu}^{(1)}(M)
  =
  -g_{\mu\nu}
  +{(p_2+p_3)_\mu(p_2+p_3)_\nu\over M^2}.
  \label{eq:vector-projector}
\end{equation}
Here $M$ is the mass of the corresponding vector kaon resonance. 
The amplitudes of $B\to\chi_{c1}\pi K$ involving the intermediate $K^\ast(892)$ and $K^\ast(1680)$ contributions are
\begin{align}
  \mathcal A_{K^\ast(892)}
  &=
  {g_1e^{\ii\phi_1}\over \Delta_{K^\ast(892)}(s_{23})}\,
  \epsilon_\chi^\mu
  P_{\mu\nu}^{(1)}(M_{K^\ast(892)})
  (p_2-p_3)^\nu ,
  \label{eq:kstar-amp}
  \\
  \mathcal A_{K^\ast(1680)}
  &=
  {g_3e^{\ii\phi_3}\over \Delta_{K^\ast(1680)}(s_{23})}\,
  \epsilon_\chi^\mu
  P_{\mu\nu}^{(1)}(M_{K^\ast(1680)})
  (p_2-p_3)^\nu ,
  \label{eq:kstar-high-amp}
\end{align}
where the denominators are defined as
\begin{equation}
  \Delta_R(s)=s-M_R^2+\ii M_R\Gamma_R .
  \label{eq:bw-convention}
\end{equation}
Here $R$ labels the specific kaon resonance, and $M_R$ and $\Gamma_R$ are its
mass and width.

For the tensor $K_2^\ast(1430)$, we introduce
\begin{equation}
  \tilde g_{\mu\nu}^{K_2^\ast}
  =
  -g_{\mu\nu}
  +{(p_2+p_3)_\mu(p_2+p_3)_\nu\over M_{K_2^\ast}^2},
\end{equation}
and
\begin{equation}
  P_{\mu\nu,\alpha\beta}^{(2)}
  =
  {1\over 2}
  \left(
  \tilde g_{\mu\alpha}^{K_2^\ast}\tilde g_{\nu\beta}^{K_2^\ast}
  +\tilde g_{\mu\beta}^{K_2^\ast}\tilde g_{\nu\alpha}^{K_2^\ast}
  \right)
  -{1\over 3}\tilde g_{\mu\nu}^{K_2^\ast}
  \tilde g_{\alpha\beta}^{K_2^\ast}.
  \label{eq:tensor-projector}
\end{equation}
The corresponding $K_2^\ast(1430)$ amplitude is
\begin{equation}
  \mathcal A_{K_2^\ast}
  =
  {g_2e^{\ii\phi_2}\over \Delta_{K_2^\ast}(s_{23})}\,
  \epsilon_\chi^\alpha (P-2p_1)^\beta
  P_{\mu\nu,\alpha\beta}^{(2)}
  p_2^\mu p_3^\nu ,
  \label{eq:k2star-amp}
\end{equation}
which corresponds to the lowest-derivative
$B\to\chi_{c1}K_2^\ast$ production vertex retained here.
The contribution from the intermediate kaon resonances retained in the following analyses is
\begin{equation}
  \mathcal A_{K^{\ast\ast}}
  =
  \mathcal A_{K^\ast(892)}
  +\mathcal A_{K_2^\ast}
  +\mathcal A_{K^\ast(1680)} .
  \label{eq:kaon-resonance-amplitude}
\end{equation}
Their masses and widths are fixed to the Particle Data Group values
\cite{ParticleDataGroup:2024cfk}, while their production couplings and phases are
fitted.

\subsection{Coupled-channel rescattering}
\label{subsec:framework-rescattering}

Besides the isovector $S$-wave $D^\ast\bar D^\ast$ channels with $J^{PC}=0^{++}$ and $2^{++}$ considered here, the $J^{PC}=1^{++}$ $D\bar D^\ast$ channel associated with the predicted $W_{c1}$ pole could in principle also contribute to the $\chi_{c1}\pi$ spectrum through coupled-channel rescattering. However, the available BaBar and Belle data show no indication of a structure near the $D\bar D^\ast$ threshold. We therefore neglect the $D\bar D^\ast$ rescattering contribution and restrict the present analysis to the $J=0$ and $J=2$ $D^\ast\bar D^\ast$ sectors.

For each $J=0,2$, the coupled-channel basis is
\begin{equation}
  a=\chi_{c1}\pi,\qquad b=D^\ast\bar D^\ast ,
  \label{eq:rescattering-basis}
\end{equation}
where $a$ is the observed hidden-charm channel, $b$ is the
$D^\ast\bar D^\ast$ channel, and
$E=m(\chi_{c1}\pi)=\sqrt{s_{12}}$ is the total energy of this two-body
subsystem.

We first write the decay contribution represented by
Fig.~\ref{fig:decay-mechanism}($c$).  If the weak vertex produces channel
$i=a,b$, the contribution that undergoes coupled-channel rescattering and
emerges in the observed channel $a$ is governed by $T_{ia}^{(J)}$.  
Then the partial-wave rescattering amplitude is
\begin{equation}
  U_J^{\rm gen}(E)
  =
  f_\Lambda[k_a(E)]
  \sum_{i=a,b} f_{Ji}\,G_i(E)\,T_{ia}^{(J)}(E),
  \quad J=0,2 .
  \label{eq:general-rescattering-source-function}
\end{equation}
Here $f_{Ji}$ is the effective weak-production coupling to channel $i$,
$G_i$ is its two-body loop function, $T_{ia}^{(J)}$ is the 
resummed coupled-channel rescattering $T$ matrix element, $k_a$ is the
$\chi_{c1}\pi$ relative momentum, and $f_\Lambda$ is the regulator
specified below.  


To assess the rescattering contribution initiated by the specified-angular-momentum source of
$\chi_{c1}\pi$ and $D^\ast\bar D^\ast$, we compare their weak-production topologies in the factorization
picture.  In the color-allowed external-$W$-emission topology, the virtual
$W$ emitted in $b\to cW^-$ produces the $\bar c s$ pair, while the charm
quark produced at the $b$ vertex combines with the spectator quark.  The
two resulting weak currents subsequently hadronize into the
$D^\ast\bar D^\ast K$ final state.  In the second
topology, the $c$ and $\bar c$ fields belonging to different currents must
be combined into a color-singlet compact charmonium $\chi_{c1}$.  This requires the
internal-emission color rearrangement and carries the color-suppressed
Wilson-coefficient combination
\cite{Bauer:1986bm,Neubert:1997uc}.  The corresponding open- and
hidden-charm production vertices may be written schematically as
\begin{align}
&\mathcal M_{\rm ext}(B\to K [D^\ast\bar D^\ast]_J)
\propto
\langle D^\ast|J_{cb}^\mu|B\rangle
\langle\bar D^\ast K|J_{\mu,sc}|0\rangle, \nonumber\\
&\mathcal M_{\rm int}(B\to[\chi_{c1}\pi]_J K)
\propto
\langle\chi_{c1}|
  \bar c\gamma^\mu\gamma_5 c
|0\rangle
\langle\pi K|J_{\mu,sb}|B\rangle ,
\label{eq:factorized-production-topologies}
\end{align}
where $J_{ij}^\mu=\bar i\gamma^\mu(1-\gamma_5)j$.  

The $\chi_{c1}\pi$ source is
further suppressed by the $P$-wave $\chi_{c1}\pi$ loop:
${\rm Im}\,G_a\propto k_a^3f_\Lambda^2(k_a)$, compared with
$k_bf_\Lambda^2(k_b)$ for an $S$-wave $D^\ast\bar D^\ast$ loop.  
Combining the above arguments, 
we retain
only the channel-$b$ source and neglect
the rescattering contribution initiated by the channel-$a$ source in the following analysis. Therefore, the partial-wave rescattering amplitude used in the fits
is
\begin{equation}
  U_J(E)
  =
  f_{Jb}\,G_b(E)\,f_\Lambda[k_a(E)]\,T_{ba}^{(J)}(E),
  \qquad J=0,2 .
  \label{eq:rescattering-source-function}
\end{equation}
The $B$-meson decay amplitudes with the $J=0$ and $J=2$ rescattering are
\begin{align}
  \mathcal A_{\rm resc}^{(0)}
  &=
  U_0(E)\,\epsilon_\chi\cdot p_2,
  \label{eq:scalar-rescattering-amp}
  \\
  \mathcal A_{\rm resc}^{(2)}
  &=
  U_2(E)\,
  \epsilon_\chi^\mu
  \Pi_{\mu\nu,\alpha\beta}^{(2)}
  p_2^\nu p_3^\alpha p_3^\beta .
  \label{eq:tensor-rescattering-amp}
\end{align}
For the tensor rescattering contribution, the spin-two projector is constructed from the
total $\chi_{c1}\pi$ momentum, i.e., 
\begin{align}
\Pi_{\mu\nu,\alpha\beta}^{(2)}
  &=
  {1\over 2}
  \left(
  \tilde g_{\mu\alpha}^{D^\ast\bar D^\ast}
  \tilde g_{\nu\beta}^{D^\ast\bar D^\ast}
  +
  \tilde g_{\mu\beta}^{D^\ast\bar D^\ast}
  \tilde g_{\nu\alpha}^{D^\ast\bar D^\ast}
  \right)   \nonumber\\
  &-{1\over 3}
  \tilde g_{\mu\nu}^{D^\ast\bar D^\ast}
  \tilde g_{\alpha\beta}^{D^\ast\bar D^\ast}, \nonumber
  \label{eq:rescattering-tensor-projector}
\end{align}
with
\begin{equation}
  \tilde g_{\mu\nu}^{D^\ast\bar D^\ast}
  =
  -g_{\mu\nu}
  +{(p_1+p_2)_\mu(p_1+p_2)_\nu
  \over (M_{D^\ast\bar D^\ast}^{\rm th})^2},
  \label{eq:rescattering-transverse-metric}
\end{equation}
where $M_{D^\ast\bar D^\ast}^{\rm th}$ is the isospin-averaged
$D^\ast\bar D^\ast$ threshold mass.

We next specify the coupled-channel $T$ matrix that enters
Eq.~\eqref{eq:general-rescattering-source-function}.  The partial-wave potential in
the $(a,b)$ basis is
\begin{equation}
  V_J
  =
  \begin{pmatrix}
    0 & a_J\\
    a_J & b_J
  \end{pmatrix},
  \qquad J=0,2 ,
  \label{eq:potential}
\end{equation}
where $a_J$ represents the coupling between the $\chi_{c1}\pi$ and $D^\ast\bar D^\ast$
channels, while $b_J$ is the elastic $D^\ast\bar D^\ast$ contact
interaction. Here, we set the $aa$ potential to zero, as the $\chi_{c1}\pi$ elastic interaction is expected to be strongly suppressed because the $\chi_{c1}$ contains no light valence quarks, in contrast to the $D^\ast\bar D^\ast$ system, where the interaction can be driven by the light quarks in the two mesons.


The coupled-channel $T$ matrix is obtained from the
Lippmann-Schwinger equation
\begin{align}
  T_J(E)
  &=
  V_J+V_JG_J(E)T_J(E),
  \label{eq:t-matrix}
\end{align}
with
\begin{equation}
  G_J(E)
  =
  \begin{pmatrix}
    G_a(E) & 0\\
    0 & G_b(E)
  \end{pmatrix}.
  \label{eq:loop-matrix}
\end{equation}
Here $G_J$ is the diagonal matrix of two-body loop functions.
All $T$ matrix elements share the same denominator,
\begin{equation}
  D_J(E)=1-b_JG_b(E)-a_J^2G_a(E)G_b(E),
  \label{eq:t-denominator}
\end{equation}
and the $T$ matrix elements are
\begin{align}
  T_{ba}^{(J)}(E)
  &=
  T_{ab}^{(J)}(E)
  ={a_J\over D_J(E)},
  \nonumber\\
  T_{aa}^{(J)}(E)
  &=
  {a_J^2G_b(E)\over D_J(E)},
  \nonumber\\
  T_{bb}^{(J)}(E)
  &=
  {b_J+a_J^2G_a(E)\over D_J(E)} .
  \label{eq:t-elements}
\end{align}
The off-diagonal element $T_{ba}^{(J)}=T_{ab}^{(J)}$ describes the transition between the $D^\ast\bar D^\ast$ channel $b$ and the $\chi_{c1}\pi$ channel $a$. Compared with the retained $T_{ba}^{(J)}$ contribution, the $T_{aa}^{(J)}$ term contains an additional loop function $G_b(E)$, providing an extra suppression factor. This further supports neglecting the $T_{aa}^{(J)}$ contribution in the present analysis. The common denominator $D_J(E)$ determines the positions of possible poles of the $T$ matrix in the complex energy plane.


For channel $i=a,b$, the nonrelativistic relative momentum is
\begin{equation}
  k_i(E)=\sqrt{2\mu_i(E-m_{{\rm th},i})},
  \label{eq:nonrel-k}
\end{equation}
with reduced mass $\mu_i$ and threshold $m_{{\rm th},i}$.  The
$\chi_{c1}\pi$ channel is treated in $P$ wave, as required by parity for
positive-parity $0^{++}$ and $2^{++}$ configurations, while the
$D^\ast\bar D^\ast$ channel is treated in $S$ wave near the
$D^\ast\bar D^\ast$ threshold.
The loop function and regulator are
\begin{align}
  G_i(E)
  &=
  \int { \dd^3q\over (2\pi)^3}
  {q^{2L_i} f_\Lambda^2(q)
  \over E-m_{{\rm th},i}-q^2/(2\mu_i)+\ii0},
  \nonumber\\
  f_\Lambda(q)
  &=
  {\Lambda^2\over \Lambda^2+q^2},
  \label{eq:loop-generic}
\end{align}
where $q$ is the loop momentum, $L_i$ is the orbital angular momentum in
channel $i$, with $L_a=1$ and $L_b=0$, and $f_\Lambda(q)$ is a
monopole regulator.  The cutoff is fixed to $\Lambda=0.5~\GeV$ in all fits
and pole searches.

With the sign convention in Eq.~\eqref{eq:loop-generic}, the imaginary part on
the physical sheet above the threshold of channel $i$ is
\begin{equation}
  {\rm Im}\,G_i^{\rm I}(E+\ii0)
  =
  -\rho_i^{(L_i)}(E),  \label{eq:phase-space-rho}
\end{equation}
with
\begin{equation}
  \qquad
  \rho_i^{(L_i)}(E)
  =
  {\mu_i k_i^{2L_i+1}(E) f_\Lambda^2[k_i(E)]\over 2\pi}. \nonumber
\end{equation}
Here $\rho_i^{(L_i)}$ is the regulated two-body phase-space factor for
channel $i$.
The Riemann sheet is labeled by $(\eta_a,\eta_b)$, where the first and second entries correspond to the $\chi_{c1}\pi$ and $D^\ast\bar D^\ast$ channels, respectively, with $\eta_i=\mathrm{I}$ for the physical sheet and $\eta_i=\mathrm{II}$ for the unphysical sheet. For channel $i$, the continuation to the unphysical sheet on the upper edge of the cut is given by
\begin{equation}
  G_i^{\rm II}(E)
  =
  G_i^{\rm I}(E)
  +2\ii\rho_i^{(L_i)}(E).
  \label{eq:sheet-continuation}
\end{equation}
The pole positions are obtained from
\begin{equation}
  D_J^{(\eta_a,\eta_b)}(E_{\rm pole})=0.
  \label{eq:framework-pole-condition}
\end{equation}

The three fit schemes are designed to examine the role of the $\chi_{c1}\pi$--$D^\ast\bar D^\ast$ coupled-channel rescattering and to compare the two isovector $S$-wave $D^\ast\bar D^\ast$ spin sectors. Scheme I includes only the effective continuum amplitude and the established intermediate kaon-resonance contributions. Schemes II and III additionally include the $\chi_{c1}\pi$--$D^\ast\bar D^\ast$ coupled-channel rescattering, with $J=0$ and $J=2$, respectively. The three schemes are defined as follows:
\begin{align}
  \mathcal A_{\rm I}
  &=
  \mathcal A_{\rm cont}
  +\mathcal A_{K^{\ast\ast}},
  \nonumber\\
  \mathcal A_{\rm II}
  &=
  \mathcal A_{\rm I}
  +\mathcal A_{\rm resc}^{(0)},
  \nonumber\\
  \mathcal A_{\rm III}
  &=
  \mathcal A_{\rm I}
  +\mathcal A_{\rm resc}^{(2)} .
  \label{eq:scheme-amplitudes}
\end{align}
This comparison allows us to examine whether coupled-channel rescattering improves the description of the $m(\chi_{c1}\pi)$ spectrum near the $D^\ast\bar D^\ast$ threshold and whether the data favor $J=0$ or $J=2$. No molecular state is introduced a priori in the fits, and possible molecular poles are identified only through analytic continuation of the fitted $T_J$ matrices.

The theoretical invariant-mass spectra of $B\to\chi_{c1}\pi K$ are obtained from the full amplitudes
in Eq.~\eqref{eq:scheme-amplitudes}.  For
$X={\rm I},{\rm II},{\rm III}$, we denote the spin-summed squared amplitude
by
\begin{equation}
  \overline{|\mathcal A_X|^2}
  \equiv
  \sum_{r_\chi}
  \left|
  \mathcal A_X(p_1,p_2,p_3,\epsilon_\chi(r_\chi))
  \right|^2 ,
  \label{eq:amp-squared}
\end{equation}
where $r_\chi$ labels the $\chi_{c1}$ polarization state. Then
the corresponding $\chi_{c1}\pi$ spectrum is
\begin{equation}
  { \dd\Gamma_X\over \dd m(\chi_{c1}\pi)}
  =
  {2\sqrt{s_{12}}\over (2\pi)^3\,32m_B^3}
  \int_{s_{23}^-}^{s_{23}^+}
  \dd s_{23}\,
  \overline{|\mathcal A_X|^2},
  \label{eq:mass-spectrum}
\end{equation}
in which $s_{12}=m^2(\chi_{c1}\pi)$, and $\Gamma_X$ denotes the partial decay width of $B\to\chi_{c1}\pi K$ calculated with
$\mathcal A_X$. With the Kall\'en function
\begin{equation}
  \lambda(x,y,z)=x^2+y^2+z^2-2xy-2xz-2yz,
\end{equation}
the kinematic integration limits are
\begin{align}
  s_{23}^{\pm}
  &=
  m_\pi^2+m_K^2
  +
  \Big[
  (s_{12}+m_\pi^2-m_{\chi_{c1}}^2)
  (m_B^2-s_{12}-m_K^2)
  \nonumber\\
  &\hspace{0.4cm}
  \pm
  \lambda^{1/2}(s_{12},m_{\chi_{c1}}^2,m_\pi^2)
  \lambda^{1/2}(m_B^2,s_{12},m_K^2)
  \Big]{1\over 2s_{12}},
  \label{eq:s23-boundaries}
\end{align}
where $s_{23}^{\pm}$ are the kinematic limits of $s_{23}$ at fixed
$s_{12}$.
The $K\pi$ spectrum is obtained analogously by fixing
$s_{23}=m^2(K\pi)$ and integrating over the kinematically allowed
$s_{12}$ range.  


Within each fit scheme, a common set of dynamical parameters describes the correlated $\chi_{c1}\pi$ and $K\pi$ spectra. The resulting invariant-mass spectra are discussed next, and the fitted amplitudes are subsequently used to predict the angular distributions in Sec.~\ref{sec:angular}.

\section{Invariant-mass spectra and pole analysis}
\label{sec:invariant-mass}

\subsection{Fit setup}
\label{subsec:fitting-strategy}

We fit the $m(\chi_{c1}\pi)$ and $m(K\pi)$ distributions from the BaBar and Belle $B\to\chi_{c1}\pi K$ data~\cite{BaBar:2011hrz,Belle:2015opn}. For BaBar, we use the combined $\chi_{c1}\pi^+$ spectrum from $B^0\to\chi_{c1}K^-\pi^+$ and $B^+\to\chi_{c1}K_S^0\pi^+$, together with the corresponding $K^-\pi^+$ and $K_S^0\pi^+$ spectra. For Belle, we use the $m(\chi_{c1}\pi^-)$ and $m(K^+\pi^-)$ distributions from $B^0\to\chi_{c1}K^+\pi^-$. The BaBar and Belle data are fitted independently.

For each collaboration's data set, all three schemes defined in Eq.~\eqref{eq:scheme-amplitudes} are fitted independently, with the effective continuum coupling and the kaon-resonance production strengths and phases reoptimized in each case. In the fit plots of the invariant-mass spectra presented below, we denote the total result of Scheme I without coupled-channel rescattering by the gray curves, while the red curves represent the total results of Schemes II and III including the $J=0$ and $J=2$ rescattering contributions, respectively.

The fitted quantities are
the effective continuum coupling, the production strengths and phases of the intermediate
kaon-resonance contributions, the production coupling $f_{Jb}$ in Scheme
II or III, and the corresponding partial-wave potential parameters $a_J$ and
$b_J$.  It is worth emphasizing that the theoretical calculation gives differential decay rates of $B\to\chi_{c1}\pi K$, whereas the
experimental invariant-mass distributions are reported as event yields.  Their overall
normalization difference can be absorbed into the fitted production coupling constants.
The fitted parameters are collected in Table~\ref{tab:fitted-parameters}.

The goodness of fit is evaluated by minimizing $\chi^2$, defined from the experimental event-yield points as
\begin{equation}
  \chi^2
  =
  \sum_{\alpha,n}
  { \left(N_{\alpha n}^{\rm exp}
  -N_{\alpha n}^{\rm th}\right)^2
  \over \sigma_{\alpha n}^2 },
  \label{eq:chi2-definition}
\end{equation}
where $\alpha$ labels the two-body invariant-mass spectrum being fitted and $n$ labels a displayed
data point.  $N_{\alpha n}^{\rm exp}$ and $\sigma_{\alpha n}$ are the
experimental event yield and uncertainty, while $N_{\alpha n}^{\rm th}$ is the
predicted event yield obtained from the theoretical line shape at the
corresponding invariant mass, with its normalization fixed by the production
couplings.  The number of degrees of freedom (d.o.f.) is
$N_{\rm pt}-N_{\rm par}$,
where $N_{\rm pt}$ and $N_{\rm par}$ are the total numbers of fitted data
points and free parameters, respectively.

\begin{table*}[t]
\caption{
Best-fit parameters for the BaBar and Belle invariant-mass distributions.
The three schemes are defined in Sec.~\ref{subsec:framework-rescattering}.
Units are given in the first column, phases are in radians, and uncertainties
are one standard deviation; ``$\cdots$'' denotes a parameter absent from a
scheme.
}
\label{tab:fitted-parameters}
\centering
\begingroup
\scriptsize
\setlength{\tabcolsep}{2.5pt}
\renewcommand{\arraystretch}{1.35}
\begin{ruledtabular}
\begin{tabular}{l ccc|ccc}
Parameter
& \shortstack{Scheme I\\(BaBar)}
& \shortstack{Scheme II\\(BaBar)}
& \shortstack{Scheme III\\(BaBar)}
& \shortstack{Scheme I\\(Belle)}
& \shortstack{Scheme II\\(Belle)}
& \shortstack{Scheme III\\(Belle)} \\
\colrule
$g_0$ $(\GeV^{-1})$
& $2.99\pm0.16$
& $3.25\pm0.11$
& $3.66\pm0.06$
& $5.15\pm0.24$
& $4.46\pm0.29$
& $5.09\pm0.09$ \\
$g_1$ $(\GeV)$
& $0.41\pm0.01$
& $0.34\pm0.01$
& $0.35\pm0.01$
& $0.49\pm0.02$
& $0.51\pm0.01$
& $0.51\pm0.01$ \\
$g_2$ $(\GeV^{-1})$
& $0.70\pm0.05$
& $0.66\pm0.05$
& $0.60\pm0.03$
& $1.13\pm0.07$
& $1.24\pm0.07$
& $1.09\pm0.04$ \\
$g_3$ $(\GeV)$
& $0.68\pm0.11$
& $1.58\pm0.06$
& $1.72\pm0.05$
& $2.67\pm0.10$
& $2.23\pm0.14$
& $2.88\pm0.07$ \\
$f_{0b}$ $(\GeV^{0})$
& $\cdots$
& $12.66\pm1.81$
& $\cdots$
& $\cdots$
& $41.40\pm4.39$
& $\cdots$ \\
$a_0$ $(\GeV^{-3})$
& $\cdots$
& $149.28\pm19.76$
& $\cdots$
& $\cdots$
& $108.63\pm13.18$
& $\cdots$ \\
$b_0$ $(\GeV^{-2})$
& $\cdots$
& $-234.63\pm66.48$
& $\cdots$
& $\cdots$
& $-117.68\pm43.59$
& $\cdots$ \\
$f_{2b}$ $(\GeV^{-2})$
& $\cdots$
& $\cdots$
& $-22.06\pm3.31$
& $\cdots$
& $\cdots$
& $187.12\pm16.51$ \\
$a_2$ $(\GeV^{-3})$
& $\cdots$
& $\cdots$
& $-128.07\pm18.83$
& $\cdots$
& $\cdots$
& $24.54\pm2.17$ \\
$b_2$ $(\GeV^{-2})$
& $\cdots$
& $\cdots$
& $-216.86\pm41.31$
& $\cdots$
& $\cdots$
& $-208.85\pm16.88$ \\
$\phi_1$ (rad)
& $1.35\pm0.12$
& $5.70\pm0.13$
& $5.83\pm0.07$
& $5.37\pm0.14$
& $5.20\pm0.16$
& $5.25\pm0.08$ \\
$\phi_2$ (rad)
& $0.29\pm0.18$
& $4.15\pm0.14$
& $4.21\pm0.11$
& $4.48\pm0.11$
& $4.32\pm0.11$
& $4.47\pm0.08$ \\
$\phi_3$ (rad)
& $2.76\pm0.12$
& $1.87\pm0.08$
& $2.01\pm0.04$
& $2.29\pm0.07$
& $2.18\pm0.09$
& $2.26\pm0.06$ \\
\colrule
$\chi^2/{\rm d.o.f.}$
& $1.15$
& $0.90$
& $0.97$
& $2.24$
& $1.69$
& $2.16$ \\
\end{tabular}
\end{ruledtabular}
\endgroup
\end{table*}

\subsection{BaBar mass spectra}
\label{subsec:babar-mass}

\begin{figure*}[t]
\centering
\begingroup
\setlength{\tabcolsep}{1pt}
\renewcommand{\arraystretch}{0.88}
\begin{tabular}{ccc}
\includegraphics[width=0.333\textwidth]{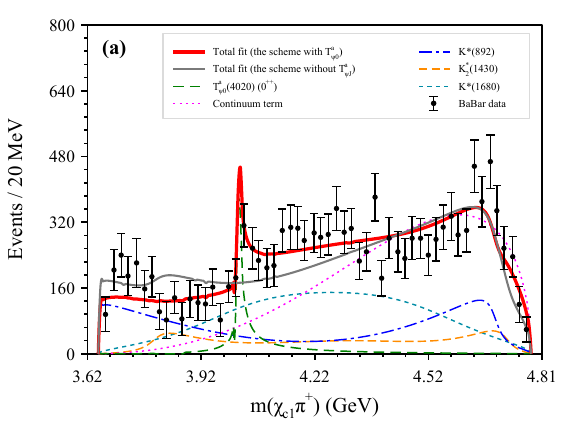} &
\includegraphics[width=0.333\textwidth]{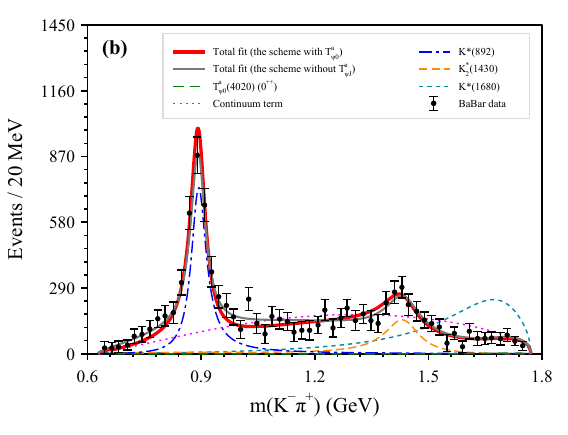} &
\includegraphics[width=0.333\textwidth]{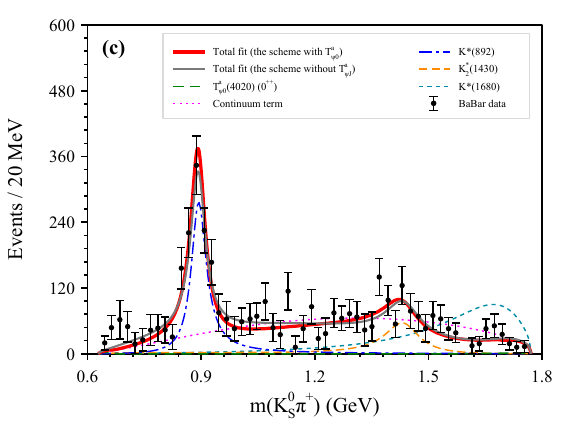} \\[-8pt]
\includegraphics[width=0.333\textwidth]{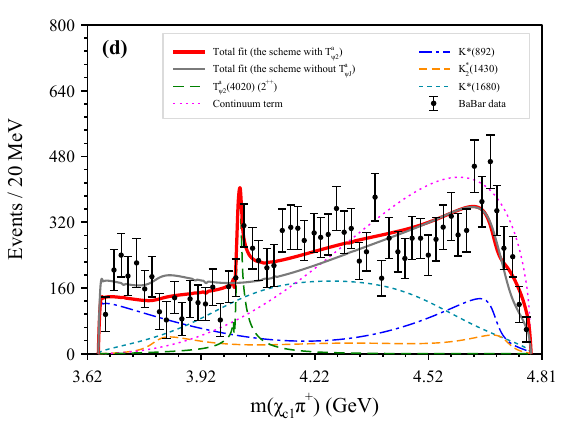} &
\includegraphics[width=0.333\textwidth]{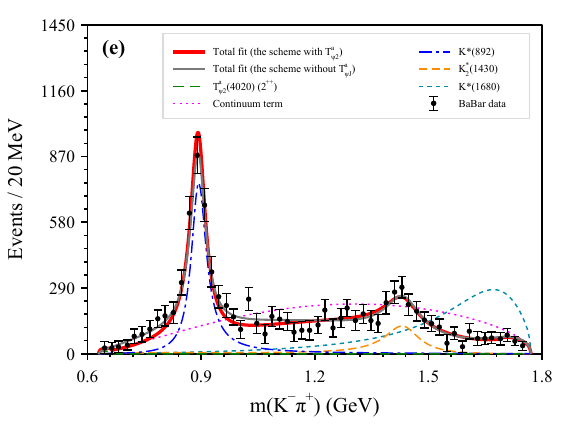} &
\includegraphics[width=0.333\textwidth]{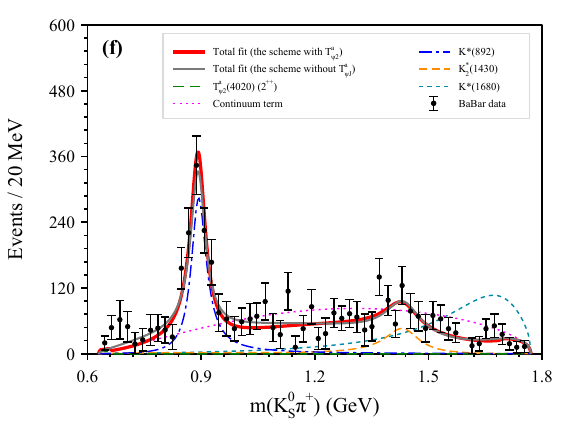}
\end{tabular}
\endgroup
\caption{
BaBar invariant-mass fits for
$B^0\to \chi_{c1}K^-\pi^+$ and
$B^+\to \chi_{c1}K_S^0\pi^+$.  The upper and lower rows show Schemes II
and III, respectively; the columns show the combined $\chi_{c1}\pi^+$,
$K^-\pi^+$, and $K_S^0\pi^+$ distributions.  Red and gray curves are
the coherent totals for the indicated scheme and Scheme I, respectively.  The remaining
curves show individual amplitude contributions without interference for the indicated scheme.
}
\label{fig:babar-mass-spectra}
\end{figure*}

Figure~\ref{fig:babar-mass-spectra} presents the simultaneous fits to the BaBar $\chi_{c1}\pi$ and $K\pi$ invariant-mass spectra in Schemes II and III, with the total result of Scheme I shown by the gray curves for comparison. The reduced $\chi^2$ values are $\chi^2/{\rm d.o.f.}=1.15,0.90,0.97$ for Schemes I, II, and III, respectively. In the two $K\pi$ spectra, the prominent $K^\ast(892)$ peak is clearly reproduced, while the broad $K_2^\ast(1430)$ and $K^\ast(1680)$ contributions are visible toward the higher-mass region. The red and gray curves are nearly identical in the $K\pi$ spectra, since the reflection of the $\chi_{c1}\pi$--$D^\ast\bar D^\ast$ rescattering contribution is spread over the kinematically allowed $K\pi$ mass range and is therefore strongly diluted.

In the $\chi_{c1}\pi^+$ spectrum, the kaon resonances generate characteristic reflections over a broad mass range. The $K^\ast(892)$ reflection produces structures toward both ends of the spectrum, the $K^\ast(1680)$ gives a broad enhancement in the intermediate-mass region, and the $K_2^\ast(1430)$ contribution varies more smoothly. Together with the effective continuum, these  contributions describe most of the observed $\chi_{c1}\pi$ line shape, as shown by the gray curve for Scheme I. Around $4.02~\GeV$, however, the data still show a tendency toward an enhancement near the $D^\ast\bar D^\ast$ threshold that is not reproduced by these reflection contributions alone. Including the $J=0$ or $J=2$ $\chi_{c1}\pi$--$D^\ast\bar D^\ast$ rescattering contribution in Schemes II and III, respectively, indeed produces a pronounced threshold peak and improves the fit. The reduced $\chi^2/{\rm d.o.f.}$ values of 0.90 and 0.97 thus favor the inclusion of coupled-channel rescattering. The improvement is somewhat larger for the $0^{++}$ scheme, although the two spin assignments give qualitatively similar descriptions of the cusp-like structure near the $D^\ast\bar D^\ast$ threshold.

\subsection{Belle mass spectra}
\label{subsec:belle-mass}

\begin{figure*}[t]
\centering
\begingroup
\setlength{\tabcolsep}{1pt}
\renewcommand{\arraystretch}{0.88}
\begin{tabular}{cc}
\includegraphics[width=0.498\textwidth]{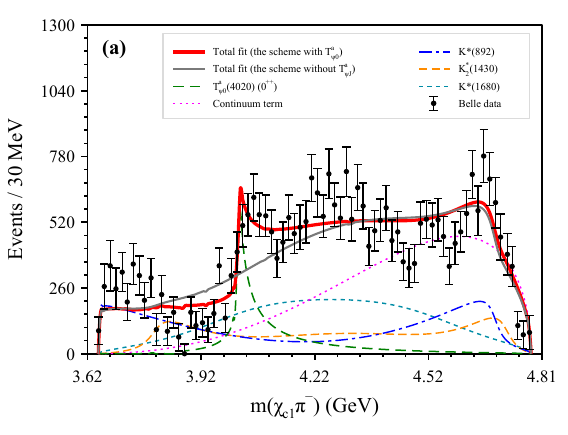} &
\includegraphics[width=0.498\textwidth]{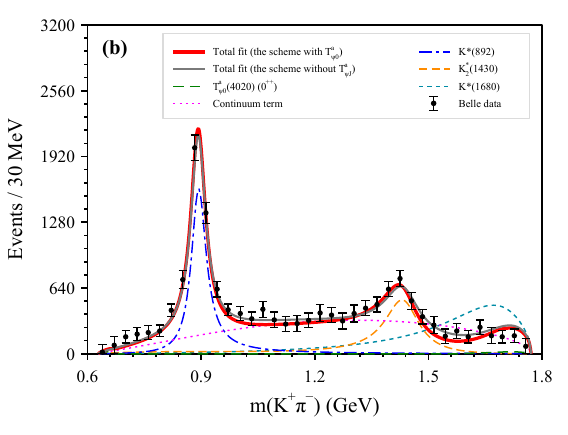} \\[-10pt]
\includegraphics[width=0.498\textwidth]{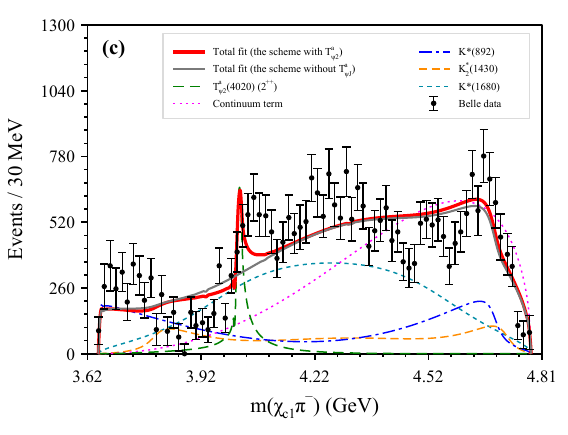} &
\includegraphics[width=0.498\textwidth]{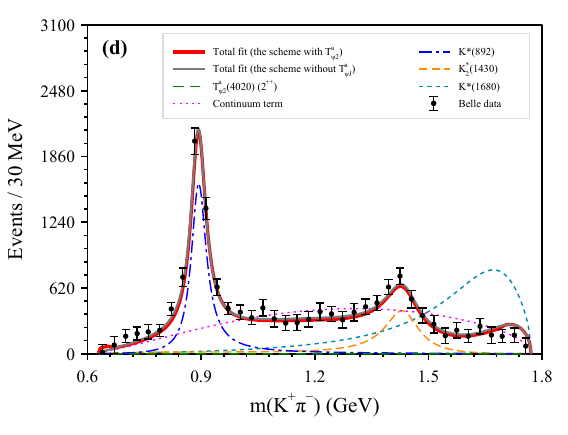}
\end{tabular}
\endgroup
\caption{
Belle invariant-mass fits for $B^0\to \chi_{c1}K^+\pi^-$.  The upper and
lower rows show Schemes II and III, respectively; the columns show the
$\chi_{c1}\pi^-$ and $K^+\pi^-$ distributions.  Red and gray curves are
the coherent totals for the indicated scheme and Scheme I, respectively.  The remaining
curves show individual amplitude contributions without interference for the indicated scheme.
}
\label{fig:belle-mass-spectra}
\end{figure*}

The Belle data fitted here are taken from the 2016 measurement of inclusive and exclusive $B$ decays to $\chi_{c1}$ and $\chi_{c2}$~\cite{Belle:2015opn}, which reported the $B^0\to\chi_{c1}K^+\pi^-$ distributions but did not perform a concrete amplitude analysis of this decay. The lower-statistics Belle data reported in 2008~\cite{Belle:2008qeq} are not included. Figure~\ref{fig:belle-mass-spectra} shows our simultaneous fits to the Belle $\chi_{c1}\pi^-$ and $K^+\pi^-$ spectra. The reduced $\chi^2$ values are $\chi^2/{\rm d.o.f.}=2.24,1.69,$ and $2.16$ for Schemes I, II, and III, respectively. The $K^+\pi^-$ line shape is similar to that in the BaBar data, with a prominent $K^\ast(892)$ peak and clear structures associated with the $K_2^\ast(1430)$ and $K^\ast(1680)$ contributions.

Interestingly, this independent Belle data set also shows that the data around and above the $D^\ast\bar D^\ast$ threshold systematically exceed the Scheme I result represented by the gray curve. In Scheme II, the $0^{++}$ rescattering contribution produces a sharp asymmetric threshold peak with an extended high-mass tail, thereby improving the description not only at threshold but also over the broader region above it. The reduced $\chi^2/{\rm d.o.f.}$ is consequently lowered from $2.24$ to $1.69$. In Scheme III, the $2^{++}$ contribution instead produces a more localized threshold peak that falls off rapidly above threshold. It therefore improves the description mainly in the immediate vicinity of the $D^\ast\bar D^\ast$ threshold, while the description at higher masses remains essentially the same as in Scheme I. Accordingly, $\chi^2/{\rm d.o.f.}$ decreases only modestly to $2.16$. Within the present model, the Belle spectra therefore favor the $0^{++}$ rescattering contribution, although the $2^{++}$ assignment cannot be excluded from the invariant-mass spectra alone.

The $\chi_{c1}\pi$ spectra near the $D^\ast\bar D^\ast$ threshold show somewhat different behaviors in the two experiments. The BaBar data favor a more localized excess around threshold, whereas the Belle data show an excess extending over a broader mass region above threshold. Nevertheless, both independent data sets indicate that the description is improved by including the $\chi_{c1}\pi$--$D^\ast\bar D^\ast$ coupled-channel rescattering contribution, providing strong support for the pole analysis below.

\subsection{Pole analysis}
\label{subsec:pole-analysis}

We search for energy poles on all four Riemann sheets of the fitted two-channel $T_J(E)$ matrices by analytically continuing the loop functions. In the BaBar and Belle fits for the $0^{++}$ and $2^{++}$ cases, the pole closest to the $D^\ast\bar D^\ast$ threshold is consistently found on the $({\rm II},{\rm II})$ sheet defined in Sec.~\ref{subsec:framework-rescattering}, where the $\chi_{c1}\pi$ and $D^\ast\bar D^\ast$ channels are continued to their second sheets.
Table~\ref{tab:pole-positions} specifies each pole solution by its position relative to
the $D^\ast\bar D^\ast$ threshold, $E_{\rm pole}-M_{\rm th}$, and by the
corresponding pole mass,
$M_{\rm pole}=M_{\rm th}+(E_{\rm pole}-M_{\rm th})$.
We use
the isospin-averaged threshold
\begin{equation}
  M_{\rm th}=2.009+2.009=4.018~\GeV .
\end{equation}
The pole positions are in general complex because the $\chi_{c1}\pi$
channel is open.  Their imaginary parts are numerically much smaller than the
quoted pole mass uncertainties and are neglected below.
 A pole below the
$D^\ast\bar D^\ast$ threshold on the unphysical sheet with respect to that
channel is a virtual-state pole.  The negative values
$E_{\rm pole}-M_{\rm th}<0$ in Table~\ref{tab:pole-positions} therefore cannot be read as binding energies of bound states.

The pole position uncertainties are obtained by propagating the fitted
uncertainties of the contact potential parameters.  We generate 5000 Gaussian
samples of the relevant $a_J$ and $b_J$ parameters using the central values
and errors in Table~\ref{tab:fitted-parameters}.  For each sample, the solved pole on the energy plane can be mapped onto the relative momentum $k$ to the $D^\ast\bar D^\ast$ channel, with $\operatorname{Im}k>0$ and $\operatorname{Im}k<0$ corresponding to bound- and virtual-state solutions, respectively.  We then retain the central
$68\%$ of the samples according to the ${\rm Im}\,k$ distribution and
map the selected samples to the energy plane
with
$E_{\rm pole}-M_{\rm th}=k^2/(2\mu_b)$, where $\mu_b$ is the
$D^\ast\bar D^\ast$ reduced mass. These pole samples are found to remain on the same $({\rm II},{\rm II})$ sheet throughout the selected interval. The resulting asymmetric uncertainty intervals of the pole position are
shown in Table~\ref{tab:pole-positions}.

The visible cusp-like enhancements in Figs.~\ref{fig:babar-mass-spectra}
and~\ref{fig:belle-mass-spectra} occur at the
$D^\ast\bar D^\ast$ threshold, whereas the analytically continued poles in
Table~\ref{tab:pole-positions} are below the $D^\ast\bar D^\ast$ threshold.  This is the expected behavior for a
virtual state: its pole lies below the $D^\ast\bar D^\ast$ threshold, while
the physical-axis spectrum for the opened channel can show a cusp-like structure.  Interference with the effective continuum amplitude and kaon-resonance
reflections can further distort the observed peak line shape. 


In the fits with the $J=0$
$\chi_{c1}\pi$--$D^\ast\bar D^\ast$ coupled-channel rescattering,
the BaBar data support the existence of $\TpsiZero$ and give an associated virtual-state pole
$-2.15^{+2.10}_{-6.41}~\MeV$ relative to the $D^\ast\bar D^\ast$
threshold, corresponding to $4015.85^{+2.10}_{-6.41}~\MeV$.  The Belle data 
give a deeper virtual-state pole farther below the $D^\ast\bar D^\ast$ threshold,
$-18.40^{+8.58}_{-13.70}~\MeV$, or
$3999.60^{+8.58}_{-13.70}~\MeV$.  This difference is consistent with the
two line shapes in the $\chi_{c1}\pi$ spectrum:
BaBar has a more localized rise at the $D^\ast\bar D^\ast$ threshold,
whereas Belle favors a broader enhancement extending above the threshold.  


The model-dependent calculations collected in Table~\ref{tab:theory-status} give widely different predictions for the isovector $0^{++}$ $D^\ast\bar D^\ast$ channel, including bound states, virtual state, resonance, and no molecular pole at all. Among the calculations that support the existence of a molecular state $\TpsiZero$, the majority predict a bound state~\cite{Wang:2023hpp,Ding:2020dio,Lu:2025zae,Peng:2023lfw}, which is inconsistent with the virtual-state poles extracted from our data-driven analysis for the BaBar and Belle data on $B\to\chi_{c1}\pi K$. The only virtual-state prediction listed in Table~\ref{tab:theory-status} has a pole position of $3972.1^{+30}_{-56}~\MeV$~\cite{Ji:2022uie}. Although its central value lies substantially below the $D^\ast\bar D^\ast$ threshold, the large uncertainty extends into the near-threshold region and is compatible with the virtual-state pole $3999.60^{+8.58}_{-13.70}~\MeV$ obtained from the fit to the Belle data.

In the fits with the $J=2$
$\chi_{c1}\pi$--$D^\ast\bar D^\ast$ coupled-channel rescattering,
the BaBar and Belle data imply the existence of $\TpsiTwo$ and give the associated virtual poles at
$-3.42^{+2.37}_{-4.29}~\MeV$ and
$-4.26^{+1.34}_{-1.69}~\MeV$ relative to the
$D^\ast\bar D^\ast$ threshold, corresponding to pole masses
$4014.58^{+2.37}_{-4.29}~\MeV$ and
$4013.74^{+1.34}_{-1.69}~\MeV$, respectively. 
The two fitted isovector $2^{++}$ pole positions agree within their quoted uncertainties and lie close to the virtual-state prediction of Ref.~\cite{Zhang:2024fxy}, $4010.4^{+2.2}_{-3.5}~\MeV$, obtained from the interaction constrained by the $W_{c1}$ state. 
If the threshold structure observed in the $\chi_{c1}\pi$ spectrum is indeed associated with an isovector $2^{++}$ state, the agreement with Ref.~\cite{Zhang:2024fxy} is particularly relevant in view of heavy-quark spin symmetry: the isovector $2^{++}$ $D^\ast\bar D^\ast$ and $1^{++}$ $D\bar D^\ast$ contact interactions depend on the same low-energy-constant combination $C_{1a}+C_{1b}$, as discussed in Sec.~\ref{sec:introduction}. In this case, the extracted $2^{++}$ pole in this data-driven analysis would also provide indirect support for the predicted virtual $W_{c1}$ state~\cite{Zhang:2024fxy}, the isovector partner of the $X(3872)$. A consistent isovector $2^{++}$ virtual-state pole at $4009\pm2~\MeV$ was independently predicted in the contact EFT analysis constrained by the $Z_c$ and $Z_{cs}$ data~\cite{Baru:2021ddn}. Determining the quantum numbers of this near-threshold $D^\ast\bar D^\ast$ state is therefore important for establishing the isovector $D^{(\ast)}\bar D^{(\ast)}$ molecular multiplet, motivating the angular-distribution analysis in the next section.

\begin{table*}[t]
\caption{
Virtual-state poles of the fitted $T$ matrices.  Energies are given relative
to the isospin-averaged $D^\ast\bar D^\ast$ threshold,
$M_{\rm th}=4018~\MeV$, and as absolute masses.  All poles lie on the
$({\rm II},{\rm II})$ sheet.  Their numerically negligible imaginary parts
are not shown.
}
\label{tab:pole-positions}
\centering
\begingroup
\setlength{\tabcolsep}{5pt}
\renewcommand{\arraystretch}{1.35}
\begin{ruledtabular}
\begin{tabular}{lccc}
Data set & $D^\ast\bar D^\ast$ sector
& $E_{\rm pole}-M_{\rm th}$ [$\MeV$]
& $M_{\rm pole}$ [$\MeV$] \\
\colrule
BaBar & $0^{++}$
& $-2.15^{+2.10}_{-6.41}$
& $4015.85^{+2.10}_{-6.41}$ \\
Belle & $0^{++}$
& $-18.40^{+8.58}_{-13.70}$
& $3999.60^{+8.58}_{-13.70}$ \\
BaBar & $2^{++}$
& $-3.42^{+2.37}_{-4.29}$
& $4014.58^{+2.37}_{-4.29}$ \\
Belle & $2^{++}$
& $-4.26^{+1.34}_{-1.69}$
& $4013.74^{+1.34}_{-1.69}$ \\
\end{tabular}
\end{ruledtabular}
\endgroup
\end{table*}

\section{Angular-distribution predictions}
\label{sec:angular}

\subsection{Angular observables}
\label{subsec:angular-definitions}

The invariant-mass analysis in Sec.~\ref{sec:invariant-mass} shows that the $0^{++}$ and $2^{++}$ rescattering schemes can give comparable descriptions of the measured mass spectra, particularly for the BaBar data. The invariant-mass spectra alone therefore provide limited discrimination between the two spin assignments. To further distinguish them, we use the corresponding best-fit amplitudes to predict angular distributions. At present, no measured angular distributions for this decay mode are available for direct comparison, so the results presented below provide predictions for future experimental analyses.

We consider two angular variables associated with the $\chi_{c1}\pi$ and $K\pi$ subsystems. The angle $\theta_{\chi_{c1}\pi}$ is defined in the $\chi_{c1}\pi$ rest frame as the angle between the $\chi_{c1}$ momentum and the direction opposite to that of the bachelor $K$. Analogously, $\theta_{K\pi}$ is defined in the $K\pi$ rest frame as the angle between the $K$ momentum and the direction opposite to that of the bachelor $\chi_{c1}$.

In a three-body decay, the angular variables are kinematically correlated with the invariant masses. For a fixed $m(\chi_{c1}\pi)$, $\cos\theta_{\chi_{c1}\pi}$ is in one-to-one correspondence with $m(K\pi)$, while for a fixed $m(K\pi)$, $\cos\theta_{K\pi}$ is in one-to-one correspondence with $m(\chi_{c1}\pi)$. This relation allows structures in one invariant-mass spectrum to be mapped into the corresponding angular distribution. Defining $s_{12}=m^2(\chi_{c1}\pi)$ and $s_{23}=m^2(K\pi)$, the relation is
\begin{equation}
  s_{23}(\cos\theta_{\chi_{c1}\pi})
  =
  m_\pi^2+m_K^2
  +2\left(
  E_\pi^\ast E_K^\ast
  -|\vec p_\pi^{\,\ast}||\vec p_K^{\,\ast}|
  \cos\theta_{\chi_{c1}\pi}
  \right),
  \label{eq:theta-chicpi-map}
\end{equation}
where the starred quantities are evaluated in the $\chi_{c1}\pi$ rest frame. Similarly, in the $K\pi$ rest frame,
\begin{equation}
  s_{12}(\cos\theta_{K\pi})
  =
m_{\chi_{c1}}^2+m_\pi^2
  +2\left(
  E_{\chi_{c1}}^\ast E_\pi^\ast
  -|\vec p_{\chi_{c1}}^{\,\ast}||\vec p_\pi^{\,\ast}|
  \cos\theta_{K\pi}
  \right).
  \label{eq:theta-kpi-map}
\end{equation}
The corresponding energies and three-momenta follow from two-body kinematics. For a particle $i$ belonging to the $ij$ subsystem, evaluated in the $ij$ rest frame,
\begin{equation}
  E_i^\ast={s_{ij}+m_i^2-m_j^2\over 2\sqrt{s_{ij}}},~
  |\vec p_i^{\,\ast}|
  =
  {\lambda^{1/2}(s_{ij},m_i^2,m_j^2)\over 2\sqrt{s_{ij}}},
  \label{eq:two-body-energy-momentum}
\end{equation}
where $j$ denotes the other particle in the subsystem. The energy and momentum of the bachelor particle $k$, evaluated in the corresponding frame, are
\begin{equation}
  E_k^\ast={m_B^2-s_{ij}-m_k^2\over 2\sqrt{s_{ij}}},~
  |\vec p_k^{\,\ast}|
  =
  {\lambda^{1/2}(m_B^2,s_{ij},m_k^2)\over 2\sqrt{s_{ij}}}.
  \label{eq:bachelor-energy-momentum}
\end{equation}
For example, a $K^{\ast\ast}$ state in the $K\pi$ subsystem is mapped into a definite range of $\cos\theta_{\chi_{c1}\pi}$ for each value of $m(\chi_{c1}\pi)$, whereas a structure in $m(\chi_{c1}\pi)$ generated by $\chi_{c1}\pi$--$D^\ast\bar D^\ast$ coupled-channel rescattering is correspondingly mapped into $\cos\theta_{K\pi}$ at fixed $m(K\pi)$. After integration over a finite invariant-mass interval, these correlations can generate nontrivial structures in the angular distributions. Their shapes reflect both the dynamical structures in the mass spectra and the explicit angular dependence associated with the spin and Lorentz structure of the decay amplitudes. The latter can provide additional sensitivity to the $0^{++}$ and $2^{++}$ assignments even when their invariant-mass spectra are similar. The angular distributions presented below are predictions based entirely on the amplitudes determined from the invariant-mass fits.

\subsection{\texorpdfstring{$\cos\theta_{\chi_{c1}\pi}$}
{cos theta chi_c1 pi} distributions}
\label{subsec:theta-chicpi}

\begin{figure*}[t]
\centering
\begingroup
\setlength{\tabcolsep}{1pt}
\renewcommand{\arraystretch}{0.86}
\begin{tabular}{ccc}
\includegraphics[width=0.333\textwidth]{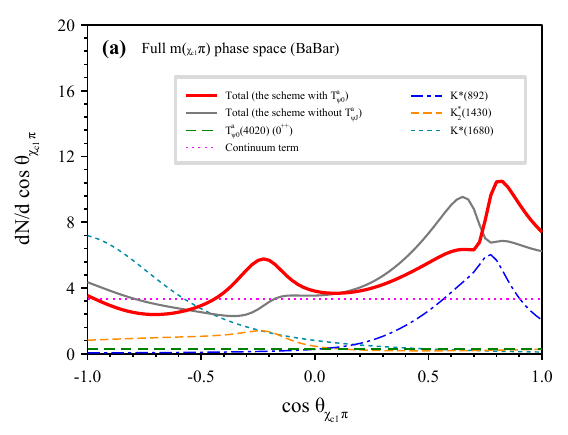} &
\includegraphics[width=0.333\textwidth]{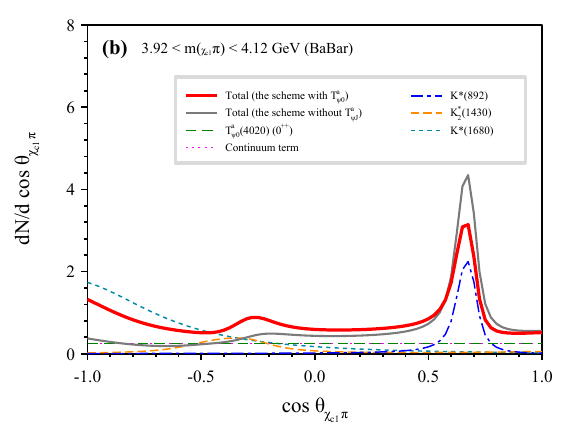} &
\includegraphics[width=0.333\textwidth]{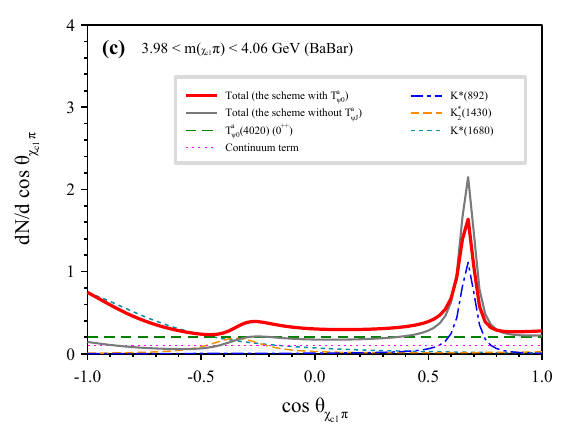} \\[-8pt]
\includegraphics[width=0.333\textwidth]{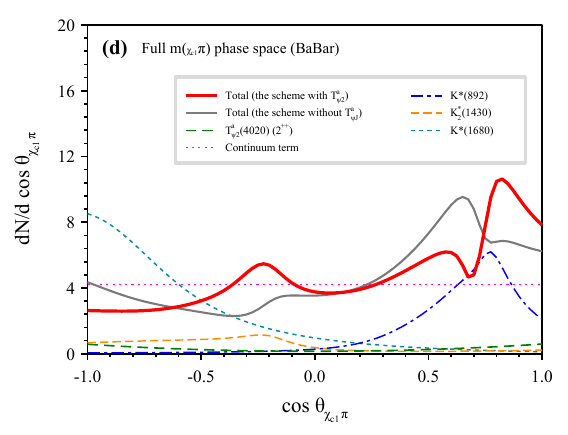} &
\includegraphics[width=0.333\textwidth]{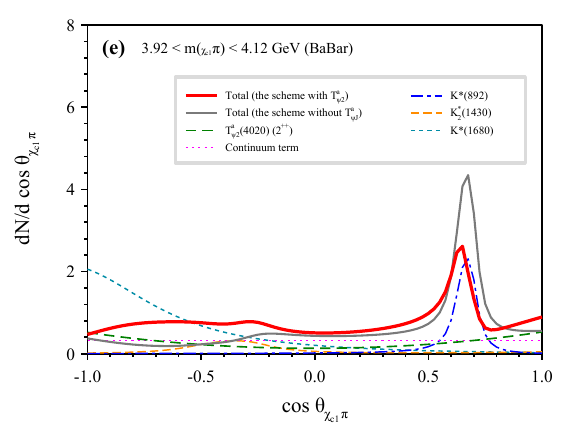} &
\includegraphics[width=0.333\textwidth]{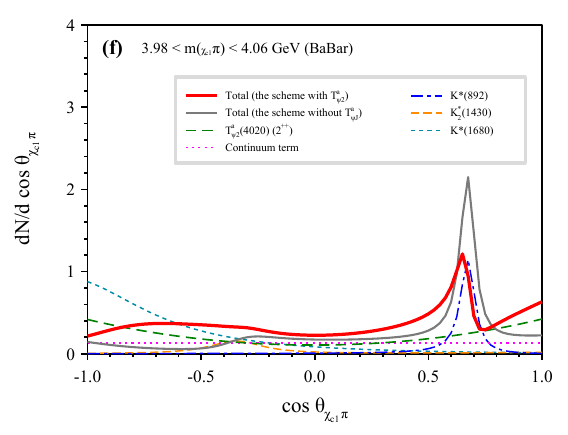} \\[-8pt]
\includegraphics[width=0.333\textwidth]{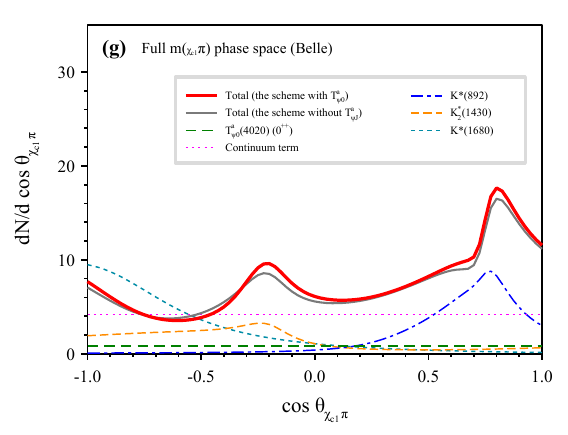} &
\includegraphics[width=0.333\textwidth]{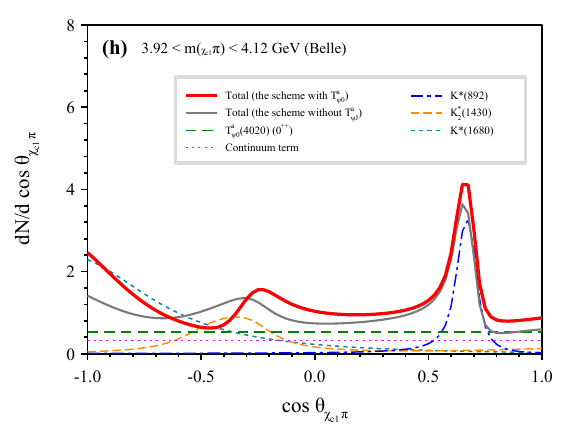} &
\includegraphics[width=0.333\textwidth]{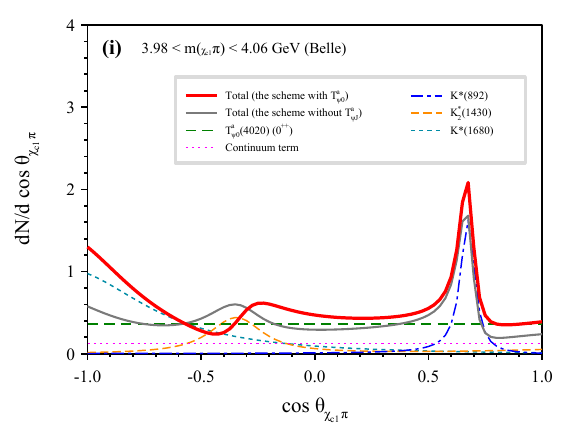} \\[-8pt]
\includegraphics[width=0.333\textwidth]{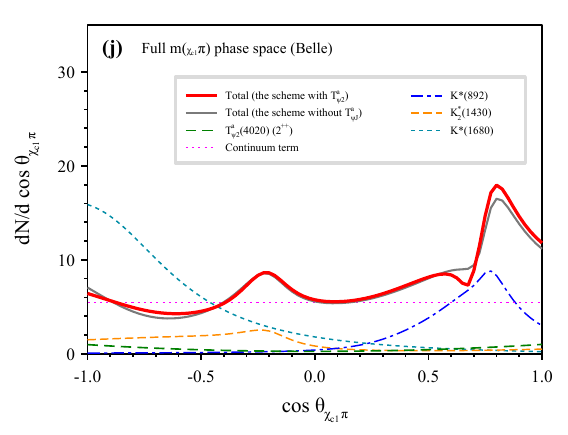} &
\includegraphics[width=0.333\textwidth]{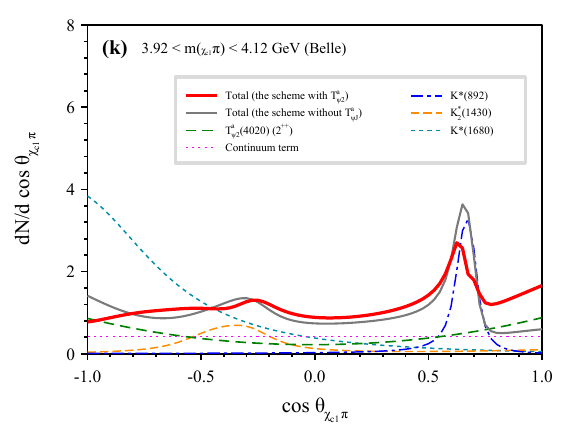} &
\includegraphics[width=0.333\textwidth]{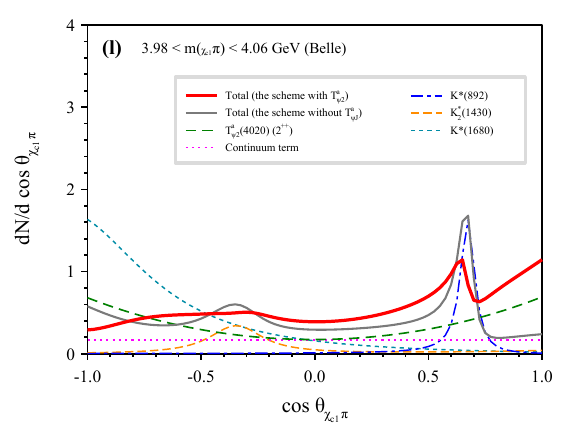}
\end{tabular}
\endgroup
\caption{
Predicted angular distributions $\dd N/\dd\cos\theta_{\chi_{c1}\pi}$.
Rows show the BaBar Scheme II, BaBar Scheme III, Belle Scheme II, and Belle
Scheme III results.  Columns cover the full $m(\chi_{c1}\pi)$ range,
$3.92$--$4.12~\GeV$, and $3.98$--$4.06~\GeV$.  Red and gray curves
are the coherent totals for the indicated scheme and Scheme I, respectively, and the remaining
curves show individual amplitude contributions without interference for the indicated scheme.
}
\label{fig:theta-chicpi}
\end{figure*}

Figure~\ref{fig:theta-chicpi} presents the predicted
$\dd N/\dd\cos\theta_{\chi_{c1}\pi}$ distributions calculated with the
amplitudes obtained from the fits to the invariant-mass spectra.  Besides the distribution
integrated over the full $m(\chi_{c1}\pi)$ range, we consider
$3.92<m(\chi_{c1}\pi)<4.12~\GeV$ and
$3.98<m(\chi_{c1}\pi)<4.06~\GeV$, both centered on the $D^\ast\bar D^\ast$ threshold.
Restricting $m(\chi_{c1}\pi)$ to the threshold region increases the relative
prominence of the rescattering contribution in the angular distribution and
hence improves the discrimination between the $0^{++}$ and $2^{++}$
assignments.  The two different window widths are used to examine how this
discrimination changes as the selected region around the threshold is
narrowed.

As discussed in Sec.~\ref{subsec:angular-definitions}, the kinematic
correlation between $m(K\pi)$ and $\cos\theta_{\chi_{c1}\pi}$ allows the
intermediate kaon resonances to produce localized enhancements in the
$\cos\theta_{\chi_{c1}\pi}$ distribution.  The calculated $K^\ast(1680)$
contribution is largest at negative $\cos\theta_{\chi_{c1}\pi}$, and the
$K_2^\ast(1430)$ contribution has a maximum around $-0.3$.  The most
prominent structure is the $K^\ast(892)$ peak at positive
$\cos\theta_{\chi_{c1}\pi}$.  It forms a relatively broad enhancement around
$0.75$--$0.8$ after integration over the full $m(\chi_{c1}\pi)$ range.
When $m(\chi_{c1}\pi)$ is restricted to either $3.92$--$4.12~\GeV$ or
$3.98$--$4.06~\GeV$, the $K^\ast(892)$ contribution is mapped into a
narrower peak around $0.65$--$0.7$, as follows from
Eq.~\eqref{eq:theta-chicpi-map}.  
For the separate rescattering contributions, the $J=0$ distribution is
independent of $\cos\theta_{\chi_{c1}\pi}$, whereas the $J=2$ distribution
has the standard symmetric tensor shape, with maxima near the two endpoints.

The Scheme II amplitude, which contains the $J^{PC}=0^{++}$ rescattering
contribution, predicts a total angular distribution that decreases from the
$K^\ast(892)$ peak and then remains approximately flat toward
$\cos\theta_{\chi_{c1}\pi}=1$ for the selected $m(\chi_{c1}\pi)$ window.  By contrast, the Scheme III amplitude, which contains the
$J^{PC}=2^{++}$ rescattering contribution, gives a total angular
distribution that reaches a local minimum at
$\cos\theta_{\chi_{c1}\pi}$ values larger than the position of the
$K^\ast(892)$ peak and then rises rapidly toward the positive endpoint at
$\cos\theta_{\chi_{c1}\pi}=1$.  The pronounced rise toward the positive
endpoint is the $2^{++}$ characteristic prediction.  This difference is most visible in
the narrower $m(\chi_{c1}\pi)$ window, which enhances the rescattering
contribution while confining the prominent $K^\ast(892)$ contribution to a
narrower angular region.

It is worth emphasizing that the same contrast is predicted by the amplitudes determined independently
from the BaBar and Belle mass spectra fits.  That two independent data sets lead to the
same spin-dependent shapes in the $\cos\theta_{\chi_{c1}\pi}$ distribution makes the
rapid rise toward the
$\cos\theta_{\chi_{c1}\pi}=1$ a strong
signature for identifying the $2^{++}$ assignment within the present amplitude
framework.

The rapid positive-endpoint rise predicted by the Scheme III amplitude also
differs clearly from the Scheme I result.  By contrast, the Scheme II
amplitude gives an angular distribution shape closer to Scheme I.  Therefore, observing the rise toward
$\cos\theta_{\chi_{c1}\pi}=1$ in a narrower integrated interval $3.98<m(\chi_{c1}\pi)<4.06~\GeV$ in future measurements would effectively distinguish the $2^{++}$ rescattering
contribution, whereas an approximately flat distribution alone would not
separate the $0^{++}$ rescattering hypothesis from Scheme I.  In the latter
case, the $m(\chi_{c1}\pi)$ spectrum remains necessary for establishing the
role of the $0^{++}$ rescattering contribution.

\subsection{\texorpdfstring{$\cos\theta_{K\pi}$}
{cos theta K pi} distributions}
\label{subsec:theta-kpi}

The invariant mass fits in Sec.~\ref{sec:invariant-mass} favor the inclusion
of the $\chi_{c1}\pi$--$D^\ast\bar D^\ast$ coupled-channel rescattering
contribution. Nevertheless, the data precision of the present $m(\chi_{c1}\pi)$ spectra is
insufficient to claim the establishment of either a $\TpsiZero$ or a $\TpsiTwo$.  Both spin
hypotheses produce a narrow cusp-like enhancement at the
$D^\ast\bar D^\ast$ threshold, whose direct measurement in the
invariant-mass spectrum generally requires sufficiently fine
mass bins. Therefore, we here propose a complementary
way to search for this $T_{\psi J}^a$ signal by measuring the $\cos\theta_{K\pi}$ distribution.  When events are integrated over a finite
$K\pi$ mass window, the kinematic correlation maps the narrow
threshold cusp into a broader structure in the $\cos\theta_{K\pi}$ distribution.  The signal can therefore be
tested without resolving it within equally narrow $m(\chi_{c1}\pi)$ bins.
This is the main purpose of the predictions for the $\cos\theta_{K\pi}$ distribution, which are
not intended as the discriminator between $J=0$ and $J=2$.

\begin{figure*}[t]
\centering
\begingroup
\setlength{\tabcolsep}{0.9pt}
\renewcommand{\arraystretch}{0.86}
\begin{tabular}{ccc}
\includegraphics[width=0.333\textwidth]{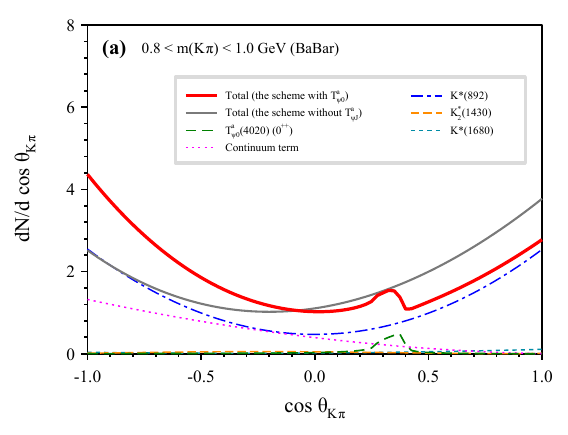} &
\includegraphics[width=0.333\textwidth]{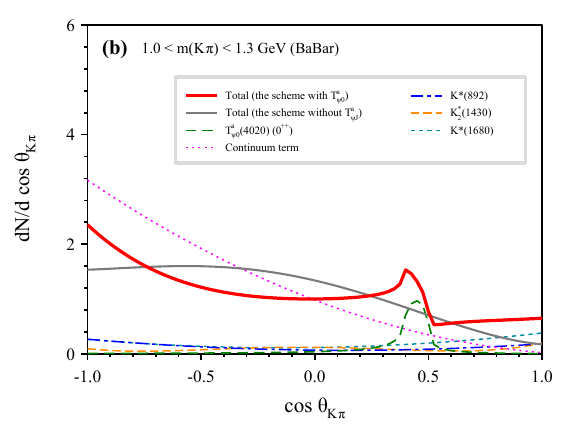} &
\includegraphics[width=0.333\textwidth]{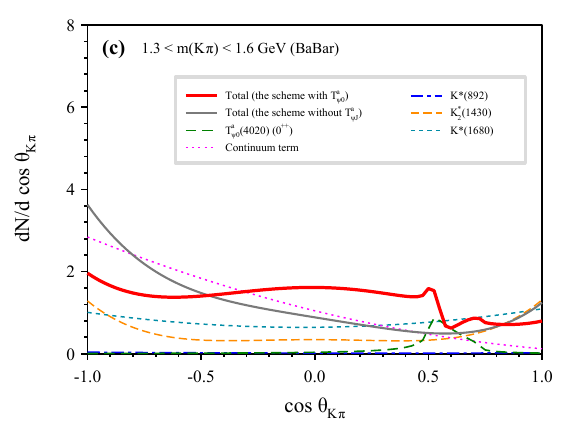} \\[-8pt]
\includegraphics[width=0.333\textwidth]{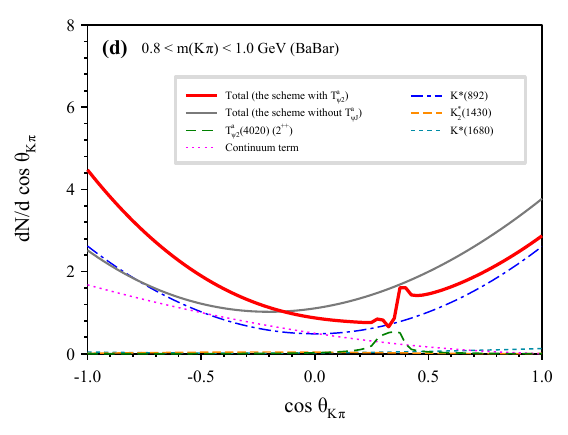} &
\includegraphics[width=0.333\textwidth]{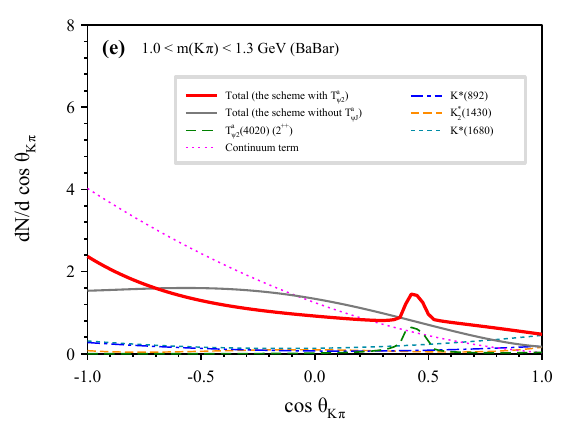} &
\includegraphics[width=0.333\textwidth]{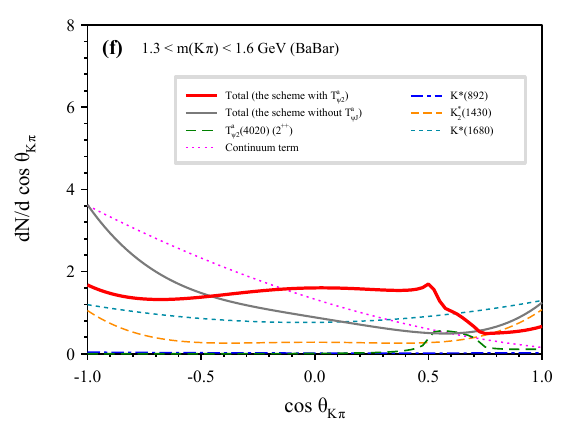} \\[-8pt]
\includegraphics[width=0.333\textwidth]{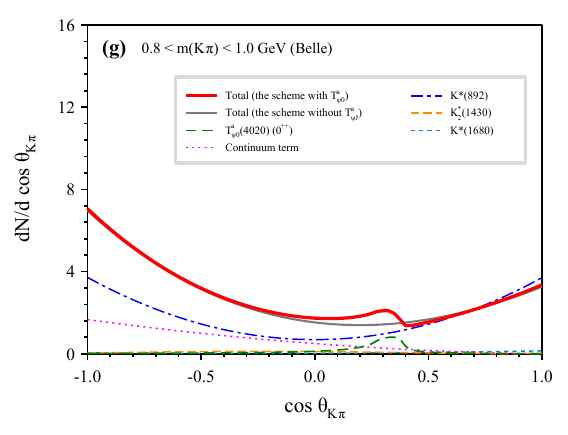} &
\includegraphics[width=0.333\textwidth]{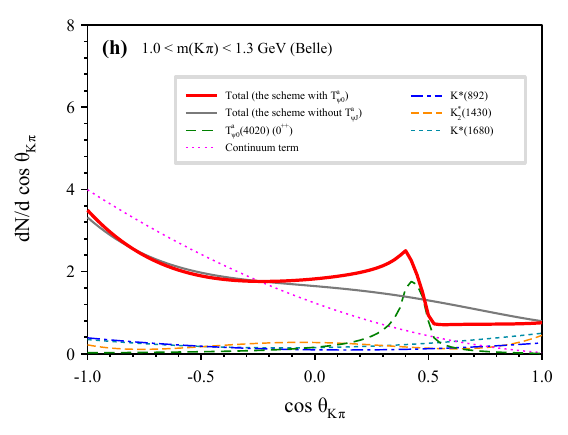} &
\includegraphics[width=0.333\textwidth]{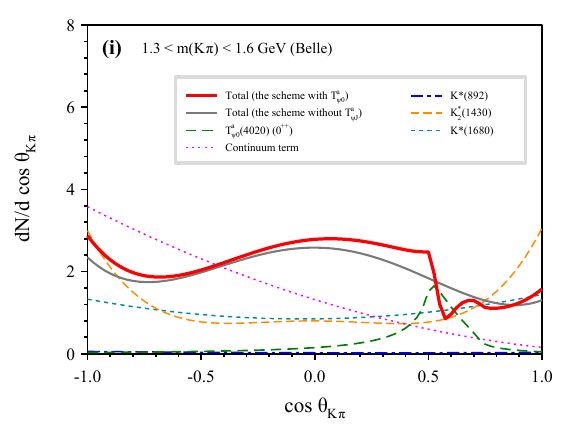} \\[-8pt]
\includegraphics[width=0.333\textwidth]{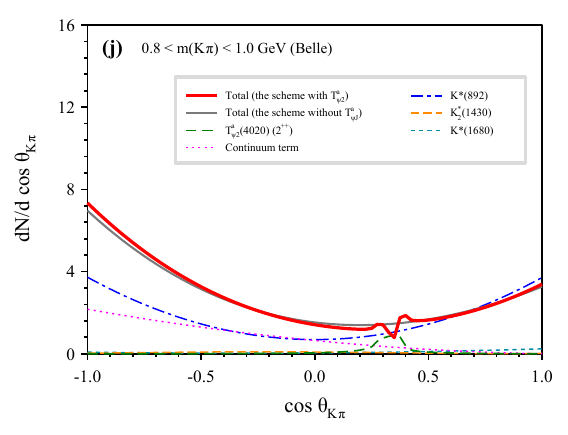} &
\includegraphics[width=0.333\textwidth]{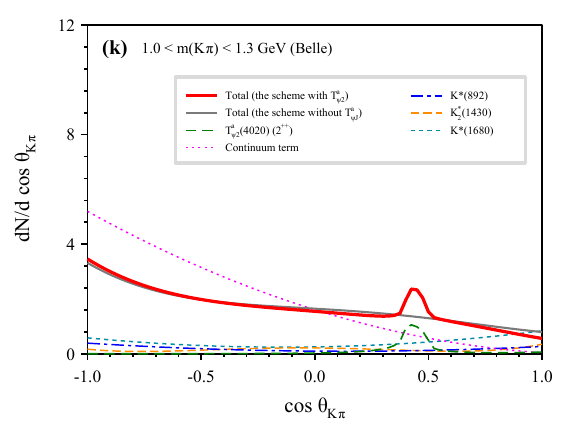} &
\includegraphics[width=0.333\textwidth]{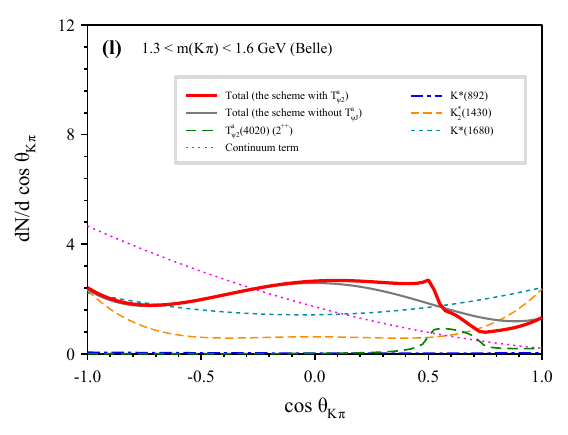}
\end{tabular}
\endgroup
\caption{
Predicted angular distributions $\dd N/\dd\cos\theta_{K\pi}$.
Rows show the BaBar Scheme II, BaBar Scheme III, Belle Scheme II, and Belle
Scheme III results.  Columns cover
$0.8<m(K\pi)<1.0~\GeV$, $1.0<m(K\pi)<1.3~\GeV$, and
$1.3<m(K\pi)<1.6~\GeV$.  Red and gray curves are the coherent totals for
the indicated scheme and Scheme I, respectively, and the remaining curves show individual
amplitude contributions without interference for the indicated scheme.
}
\label{fig:theta-kpi}
\end{figure*}

Figure~\ref{fig:theta-kpi} shows the predicted total angular distributions
$\dd N/\dd\cos\theta_{K\pi}$ integrated over
$0.8<m(K\pi)<1.0~\GeV$, $1.0<m(K\pi)<1.3~\GeV$, and
$1.3<m(K\pi)<1.6~\GeV$.  At fixed $m(K\pi)$,
Eq.~\eqref{eq:theta-kpi-map} gives a one-to-one relation between
$m(\chi_{c1}\pi)$ and $\cos\theta_{K\pi}$.  Denoting the squared
$D^\ast\bar D^\ast$ threshold mass by
$s_{\rm th}=(M_{D^\ast\bar D^\ast}^{\rm th})^2$, the threshold is mapped to
\begin{equation}
  \cos\theta_{K\pi}^{\rm th}(s_{23})
  =
  {m_{\chi_{c1}}^2+m_\pi^2
  +2E_{\chi_{c1}}^\ast E_\pi^\ast-s_{\rm th}
  \over
  2|\vec p_{\chi_{c1}}^{\,\ast}||\vec p_\pi^{\,\ast}|}.
  \label{eq:theta-kpi-threshold-map}
\end{equation}
After
integration over a finite $K\pi$ interval, the near-threshold structure is mapped into an
angular range rather than a single value of $\cos\theta_{K\pi}$.   

In the lowest interval, $0.8<m(K\pi)<1.0~\GeV$, the $K^\ast(892)$
contribution produces the dominant U-shaped dependence on
$\cos\theta_{K\pi}$.  The near-threshold structure is mapped to
$\cos\theta_{K\pi}\simeq0.3$--$0.4$.  In this region, the Scheme II
amplitude for $0^{++}$ predicts a broad local maximum followed by a decrease,
whereas the Scheme III amplitude for $2^{++}$ gives a narrow local  minimum
followed by a narrow local maximum.  The same difference occurs in the results
obtained from the BaBar and Belle fits.  In both cases, however, the local
distortion is small compared with the $K^\ast(892)$ contribution,
especially in the Belle-based prediction, so this interval is not the most
favorable one for observing the rescattering signal.

For the middle interval, $1.0<m(K\pi)<1.3~\GeV$, the angular distribution associated with $\cos\theta_{K\pi}$ gives the clearest
signal correlated with the $D^\ast\bar D^\ast$ dynamics.  The Scheme II
amplitude gives a broad enhancement extending to
$\cos\theta_{K\pi}\simeq0.4$, terminated by a sharp decrease above $0.5$.
The Scheme III amplitude gives a more localized peak around $0.4$--$0.5$.
These respective changes occur in the predictions based on the independently
fitted BaBar and Belle amplitudes.  Thus both spin hypotheses give an obvious
threshold-correlated feature in the predicted total angular distribution,
although its shape depends on the spin assignment.

In the highest interval, $1.3<m(K\pi)<1.6~\GeV$, the mapped structure
moves to $\cos\theta_{K\pi}\simeq0.5$--$0.6$.  The effective continuum,
$K_2^\ast(1430)$, and $K^\ast(1680)$ contributions are more important in
this window.  Against this background, the Scheme II amplitude gives a
local maximum followed by a sharp decrease and a recovery at larger
$\cos\theta_{K\pi}$, whereas the Scheme III amplitude gives a weaker
shoulder followed by a broader decrease.  These behaviors again occur in
the predictions from both data sets, but the larger conventional
contributions make the rescattering signal less distinct than in the middle
interval.

In addition, the characteristic
prediction here is the displacement of the mapped structure from
approximately $\cos\theta_{K\pi}=0.35$ to $0.45$ and then to $0.55$ across
the three $K\pi$ mass intervals.  Measuring this change behavior would provide a complementary strategy for establishing a narrow $\TpsiZero$ or $\TpsiTwo$ state without relying only on fine binning of the $m(\chi_{c1}\pi)$ spectrum.  Since both spin
hypotheses share this correlated displacement, the
$\cos\theta_{K\pi}$ distribution primarily tests the presence of the
near-threshold rescattering contribution.  Although the different local shapes in the $\cos\theta_{K\pi}$ distribution
predicted by Schemes II and III provide some spin sensitivity, the
$\cos\theta_{\chi_{c1}\pi}$ distribution remains
the cleaner observable for distinguishing  $0^{++}$ from $2^{++}$ for the rescattering contribution.

\section{Summary and outlook}
\label{sec:summary}

Motivated by the unsettled experimental status of the $Z_1(4050)^+$ and by
predictions of the isovector $D^{(\ast)}\bar D^{(\ast)}$ molecular
multiplets, we have studied the BaBar and Belle
$B\to\chi_{c1}\pi K$ data to search for the $S$-wave $D^\ast\bar D^\ast$ states,
$\TpsiZero$ with $J^{PC}=0^{++}$ and $\TpsiTwo$ with
$J^{PC}=2^{++}$. 
For $B\to\chi_{c1}\pi K$, we construct a unitary
$\chi_{c1}\pi$--$D^\ast\bar D^\ast$ coupled-channel rescattering amplitude and combine
it coherently with the effective continuum and intermediate-kaon-resonance
amplitudes.  The $\chi_{c1}\pi$ and $K\pi$ invariant-mass spectra are
fitted simultaneously, while the BaBar and Belle data
sets are analyzed independently.  Both data sets are better described when
the coupled-channel rescattering contribution is included.  The BaBar
spectra do not distinguish the $0^{++}$ and $2^{++}$ hypotheses for the rescattering channels, whereas
the Belle spectra favor the $0^{++}$ hypothesis within the present amplitude
framework but do not exclude $2^{++}$.

After the mass-spectra fits determine the coupled-channel amplitudes, we
analytically continue the fitted $T$ matrices to the complex energy plane to search for poles.  Under
the $0^{++}$ hypothesis, the scattering  $T$ matrices extracted from both BaBar and Belle data consistently give virtual-state poles for
$\TpsiZero$ on the $({\rm II},{\rm II})$ sheet at
$-2.15^{+2.10}_{-6.41}~\MeV$ and
$-18.40^{+8.58}_{-13.70}~\MeV$ relative to the
$D^\ast\bar D^\ast$ threshold, respectively.
Under the $2^{++}$ hypothesis, the corresponding virtual-state poles for $\TpsiTwo$ are found
at $-3.42^{+2.37}_{-4.29}~\MeV$ and
$-4.26^{+1.34}_{-1.69}~\MeV$, respectively.  The two independently determined $2^{++}$
pole positions agree within their uncertainties and are close to the
virtual state predictions in the isovector $2^{++}$ $D^\ast\bar D^\ast$ channel of
Refs.~\cite{Baru:2021ddn,Zhang:2024fxy}.

The invariant mass spectra alone do not resolve the two spin assignments.
In the narrow window $3.98<m(\chi_{c1}\pi)<4.06~\GeV$, however, the
independently fitted BaBar and Belle amplitudes predict the same qualitative
distinction in the measurable total angular distribution
$\dd N/\dd\cos\theta_{\chi_{c1}\pi}$.  At
$\cos\theta_{\chi_{c1}\pi}$ values larger than the position of the peak structure from the $K^\ast(892)$ contribution, the decay amplitude including the $0^{++}$ rescattering gives an
approximately flat distribution.  The decay amplitude for $2^{++}$
instead predicts a local minimum followed by a rapid rise toward the positive
endpoint at $\cos\theta_{\chi_{c1}\pi}=1$.  The occurrence of this
positive-endpoint rise in two independent predictions based on the BaBar and Belle data set makes it a strong
discriminator for the $2^{++}$ assignment. The other angular distribution of $\cos\theta_{K\pi}$ provides a complementary way to detect the signals of the near-threshold state $\TpsiZero$ or $\TpsiTwo$: integration over a finite $K\pi$ invariant mass interval maps
the cusp-like structure in $m(\chi_{c1}\pi)$ into a broader feature  in the $\cos\theta_{K\pi}$ distribution whose position
moves from about $\cos\theta_{K\pi}=0.35$ to about $0.55$ as the selected
$K\pi$ mass interval is increased.

Determining whether the isovector $0^{++}$ and $2^{++}$
$D^\ast\bar D^\ast$ channels contain near-threshold poles is very important
for establishing the complete isospin-spin $S$-wave $D^{(\ast)}\bar D^{(\ast)}$ molecular multiplets.  A
confirmed $0^{++}$ $\TpsiZero$ state would constrain the presently undetermined isovector scalar
interaction.  A confirmed $2^{++}$ $\TpsiTwo$ state would test the heavy quark spin symmetry relation and provide indirect
evidence for the predicted $1^{++}$ $W_{c1}$ pole, whose neutral member corresponds to an isospin partner of $X(3872)$.  A higher-statistics
data set for $B\to\chi_{c1}\pi K$ that combines the
invariant-mass spectra with the angular distributions in the selected mass
windows can test these alternatives in future Belle II and LHCb experiments.

\begin{acknowledgments}
This project was supported by the National Natural Science Foundation of China under Grants No. 12405088 and No. 12547101, and by the Start-up Funds of Chongqing University.
\end{acknowledgments}

\bibliographystyle{apsrev4-2}
\bibliography{references}

\end{document}